 \documentclass[final,5p,times,twocolumn,authoryear]{elsarticle}

\usepackage{amssymb}
\usepackage{lipsum}
\usepackage{xcolor}
\usepackage{siunitx}
\usepackage{amsmath}
\usepackage{hyperref}
\usepackage{listings}
\usepackage{xcolor}
\usepackage{adjustbox}
\usepackage[table]{xcolor}
\usepackage{colortbl}
\usepackage{aas_macros}

\usepackage{graphicx} 
\usepackage{caption}  
\usepackage[table]{xcolor}

\journal{Astronomy $\&$ Computing}

\begin{document}

\begin{frontmatter}



\title{Skykatana: a scalable framework to construct sky masks for the Vera Rubin Observatory and large astronomical surveys}

\author[icate,unsj]{Lopez, Claudio\corref{cor1}}
\ead{yoclaudioantonio@gmail.com}

\author[icate,unsj]{Donoso, Emilio}
\ead{emiliodon@gmail.com}

\author[iate,uncor]{Dominguez Romero, Mariano Javier de L.}
\ead{mariano.dominguez@unc.edu.ar}

\cortext[cor1]{Corresponding author.}

\affiliation[icate]{organization={Instituto de Ciencias Astronómicas, de la Tierra y del Espacio (ICATE), Consejo Nacional de Investigaciones Científicas y Técnicas (CONICET)},
            addressline={Av. Nazario Benavidez 8175 Oeste}, 
            city={Chimbas},
            postcode={5400}, 
            state={San Juan},
            country={Argentina}}

\affiliation[unsj]{organization={Facultad de Ciencias Exactas, Físicas y Naturales (FCEFYN-UNSJ), Universidad Nacional de San Juan},
             addressline={Av. José Ignacio de la Roza 590 Oeste},
             city={Rivadavia},
             postcode={5400},
             state={San Juan},
             country={Argentina}}

\affiliation[iate]{organization={Instituto de Astronomía Teórica y Experimental (IATE), Consejo Nacional de Investigaciones Científicas y Técnicas (CONICET)},
             addressline={Laprida 854},
             city={Córdoba},
             postcode={5000},
             state={Córdoba},
             country={Argentina}}

\affiliation[uncor]{organization={Observatorio Astronómico Córdoba
, Universidad Nacional de Córdoba},
             addressline={Laprida 854},
             city={Córdoba},
             postcode={5000},
             state={Córdoba},
             country={Argentina}}

\begin{abstract}
Modern wide-field surveys require robust spatial masks to excise bright-star halos, bleed trails, poor-quality regions, and user-defined geometry at scale. We present Skykatana, an open source pipeline that builds and combines boolean HEALPix/HEALSparse maps into science-ready masks and engineered for low-memory operation. Skykatana can efficiently construct, visualize multi-order coverage maps and generate random points in high-resolution masks over half of the celestial sphere with very limited resources and leveraging the hierarchical partition of data the HATS/LSDB framework. We demonstrate two end-to-end applications: (1) a Subaru HSC-WISE composite mask; and (2) Rubin star masks generated on demand in the Rubin Science Platform by querying HATS/LSDB Gaia data and assigning radii from empirical fits to Rubin DP1 data. We release full bright-star masks for various regions of the Rubin footprint and describe performance and scaling. The code, documentation, and examples are publicly available \href{https://github.com/samotracio/skykatana}{here}, and the LSST masks can be obtained in \href{https://osf.io/r5vw6}{this repository}.
\end{abstract}




\end{frontmatter}





\section{Introduction}
\label{introduction}
The rapid growth of modern wide-field imaging surveys has fundamentally changed the way astronomers characterize the observable Universe. Projects such as the Vera C. Rubin Observatory Legacy Survey of Space and Time (LSST; \citealt{Ivezic2019}, \citealt{PSTN019-SciPip}), the Subaru Hyper Suprime-Cam (HSC; \citealt{hsc}) survey, and the Dark Energy Survey (DES; \citealt{Porredon2021}, \citealt{Rozo2016}, \citealt{RodriguezMonroy2022}, \citealt{Rodriguez2025}) have achieved unprecedented sky coverage and depth, producing petabyte-scale datasets that must be rigorously characterized in terms of angular completeness and data quality. For most scientific applications, ranging from galaxy clustering and cosmic shear measurements to stellar population and transient studies, it is essential to know whether the survey data at every point on the sky meet the required quality criteria. This information is encoded in spatial masks that exclude regions contaminated by bright stars, diffraction spikes, detector artifacts, poor seeing, astrophysical foregrounds, or incomplete coverage, and forms the backbone of most selection functions in modern surveys.

Traditional geometric mask frameworks such as MANGLE (\citet{hamilton2004}, \citet{swanson2008}) represent the sky as unions of spherical polygons with analytic boundaries, achieving exact area computation, but requiring complex data structures, tolerance to numerical instability, and expensive spherical operations that scale poorly with the number of polygons. Pixelization-based approaches, most notably HEALPix \citep{gorski2005}, offer a simpler and faster alternative by discretizing the sphere into pixels of equal area. HEALPix is ideally suited for dense, full-sky scalar fields (e.g., maps of temperature, intensity, or depth), but becomes inefficient for sparse boolean maps because memory use scales linearly with the total number of pixels. The HEALSparse library (\citealt{hsppypi}, Dark Energy Science Collaboration) addressed this limitation by introducing a hierarchical "coverage/sparse" representation, where only the pixels actually present in the mask are stored, preserving exact geometry while reducing memory and I/O costs by orders of magnitude.

The Multi-Order Coverage framework (MOC; \citealt{fernique2015}) allows compact, hierarchical representation of arbitrary sky regions, and is widely used visualizing survey footprints, catalog coverage, and boolean region operations. The MOCpy Python library \citep{boch2019} implements an interface for manipulating MOCs in pure Python. However, MOCpy is primarily focused on region-level operations and is not optimized for high-resolution boolean masking of extended exclusion zones (e.g., around bright stars) across billions of pixels. 

The prior literature has also established methods and procedures for survey-specific bright star masking. For the Pan-STARRS 3$\pi$ survey,  (\citealt{farrow2014}) binned the survey footprint into $3.3''$ equal-area pixels, computed coverage and variance maps, and defined exclusion regions around bright stars using the UCAC4/Hipparcos catalog and a power-law fit of mask radius vs magnitude. They also applied a conservative cut on stack coverage ($\leq$3 exposures) and expanded mask borders by one pixel to mitigate edge effects. Their workflow illustrates the coupling of bright-source masks, depth/coverage flags and geometric clean-up in a modern 2-meter, wide-area imaging survey. For HSC-SSP, \citet{coupon2018} describes the construction and validation of bright-star masks based on Gaia and Tycho-2 data. Their analysis quantifies saturation effects, completeness trade-offs, and downstream impacts on number counts and clustering. These HSC masks (Arcturus release) remain as the only reference for bright-star mitigation in deep 8-meter wide-field telescope imaging. 
The angular mask applied to the galaxy samples in DES Year 3 analysis  \citep{Porredon2021, Rozo2016, RodriguezMonroy2022} imposes additional data selection cuts that affect cosmological tests, such as correlation functions, e.g., \citep{Ross2011}. More recently, DES Y6 analysis showed improved mitigation of spatially varying observational systematics through masking, as reported in \citep{Rodriguez2025}. Observational systematics encompasses effects related to observing conditions, survey strategy, and astrophysical foregrounds, such as stellar density, dust extinction, galactic cirrus, and near galaxies and globular clusters. For each of these, a map of their spatial distribution on the sky could be built up, and they are known as property maps \citep{2025arXiv250105739B}.

Therefore, new tools are needed to manage and combine the multiple property maps required to define 
the angular selection function of future surveys like the NSF-DOE Vera C. Rubin Observatory's Legacy survey of Space and Time (LSST); \cite{Ivezic2019}.

Building upon this foundation, Skykatana provides a general-purpose modular framework to construct, combine, and visualize such sparse boolean maps at survey scale. It introduces the concept of mask pipelines, composed of independent stages (e.g. footprints, bright-star masks, depth thresholds, or extended-source masks), each represented internally as a HEALSparse map. These stages can be generated from catalogs, property maps, or analytic geometric primitives, and then logically combined through union, intersection, or subtraction into a final mask suitable for scientific use. Skykatana emphasizes both computational efficiency and workflow clarity, enabling users to generate reproducible, documented masking procedures that scale from small fields to the entire Rubin footprint. Skykatana is designed to operate efficiently in distributed environments and with HATS/LSDB\footnote{\href{https://docs.lsdb.io/en/latest/}{docs.lsdb.io/en/latest/}} schemas (\citealt{caplar2025}), where astronomical catalogs are hierarchically partitioned in HEALPix cells. The code implements pixel streaming algorithms that enable processing large areas of the sky, automatically query large datasets that would be costly to search locally, and stream the resulting pixel sets directly into bit-packed maps. These features enable the generation of high-resolution masks of the entire Rubin footprint directly in the Rubin Science Platform (RSP) instance hosted at \url{https://data.lsst.cloud} \citep{omullane2024} with modest computational resources.

The new pipeline integrates advanced visualization features, MOC overlays, random point sampling, and fractional area diagnostics, as well as efficient I/O via compact FITS tables with bit-packed encoding. These capabilities enable the interactive inspection of billion-scale pixel masks within Jupyter notebooks.

In this paper, we present the design and capabilities of Skykatana, describe its internal algorithms and data model, and demonstrate two end-to-end applications: (i) a Subaru Hyper Suprime-Cam and Widefield Infrared Survey Explorer (HSC–WISE) composite mask, combining footprint, depth, and bright-star stages; (ii) a bright-star mask for the Rubin Observatory over the Wide-Fast-Deep survey area, constructed on demand via HATS/LSDB queries using empirical magnitude–radius relations derived from Rubin Data Preview 1 (DP1; \citealt{RTN095-DP1}, \citealt{dp1-lsdb}) data. 

\begingroup
The primary objectives of this work are:
\begin{enumerate}
\item to introduce Skykatana as a scalable framework for constructing and combining sky masks at the resolution and data volumes required by next-generation surveys;
\item to enable on-demand, catalog-driven mask generation through distributed query engines;
\item to provide a reproducible and composable masking logic that can be stored, combined, and reused across analyses;
\item to demonstrate these capabilities using representative masking applications drawn from existing survey data.
\end{enumerate}
\endgroup


Existing masking approaches typically fall into static, survey-specific polygon masks that are difficult to scale or adapt. Skykatana is designed to fill this gap by enabling high-resolution, dynamically generated masks that remain computationally tractable and reproducible at Rubin-scale data volumes. This gap becomes critical in modern survey workflows, where analyses are increasingly executed in cloud-based notebook environments with constrained memory, require distributed processing over billions of pixels, and demand strict reproducibility of mask definitions across teams and analyses. These practical constraints strongly motivate the design choices underlying Skykatana. The demonstration applications presented in this paper based on HSC-WISE masking and Rubin bright-star exclusion are intended to illustrate the capabilities of the Skykatana framework, rather than to provide definitive or universal scientific calibrations.

Throughout, we emphasize performance, memory efficiency, and reproducibility. Section 2 describes the pipeline architecture and algorithms; Section 3 details the calibration of the star-exclusion radii using DP1 data; Sections 4 and 5 present the two application examples described above; and Section 6 discusses limitations and future extensions.

\subsection*{Design highlights}
The Skykatana framework is designed around the following key principles:

\begin{itemize}
  \item \textbf{Scalable sparse masking}: Angular masks containing billions of HEALPix pixels are represented using sparse, bit-packed encodings and streaming I/O, enabling their use in memory-constrained notebook and cloud environments.
  \item \textbf{On-demand, catalog-driven construction}: Mask stages are generated directly from distributed catalogs via LSDB/HATS queries, avoiding the need for precomputed full-sky products.
  \item \textbf{Chunked, hierarchical workflow}: A breadth-first chunking strategy ensures bounded memory usage while scaling naturally with survey area.
  \item \textbf{Composable pipeline abstraction}: Mask stages can be combined, persisted, and reloaded reproducibly using a unified pipeline model.
  \item \textbf{Interoperability and visualization}: Native support for MOC export and interactive notebook visualization enables rapid inspection and validation of complex masks.
\end{itemize}


\section{The Skykatana framework}

\subsection{Design goals and architecture}
Skykatana approach is centered around the pipeline concept, i.e. a series of automated processing steps that transform raw data into scientifically sound information. As such, a pipeline in Skykatana provides a logical structure to create, manipulate, and store multiple masks of the sky, while also keeping a record of the parameters used in their construction. The nature and processing of the individual masks used, such as those used to isolate complex areas covered by a multiband survey under specific observing conditions, can be very different and require substantial exploratory analysis. A design goal of Skykatana is to facilitate an encapsulated but flexible object to keep track of the various effects affecting a given area of the sky, visualize geometric details, and provide the most common functionality for subsequent astronomical tasks, such as generating random points, applying the mask to arbitrary catalogs, and computing areas.

Skykatana organizes a pipeline as a series of stages, where each stage is a boolean HEALSparse map with its associated metadata. Stages are meant to describe an area of the sky that should be included or excluded from astronomical analysis for various reasons, and come from different data sources. The code allows the creation of a stage from: (i) a list of geometric primitives, (ii) a catalog of discrete sources, (iii) a HEALSparse map, and (iv) other stages. The stages can then be logically combined into a "final" mask by performing unions, subtractions, and intersections at a given target order.

\begin{figure}
    \centering
    \includegraphics[width=1\linewidth]{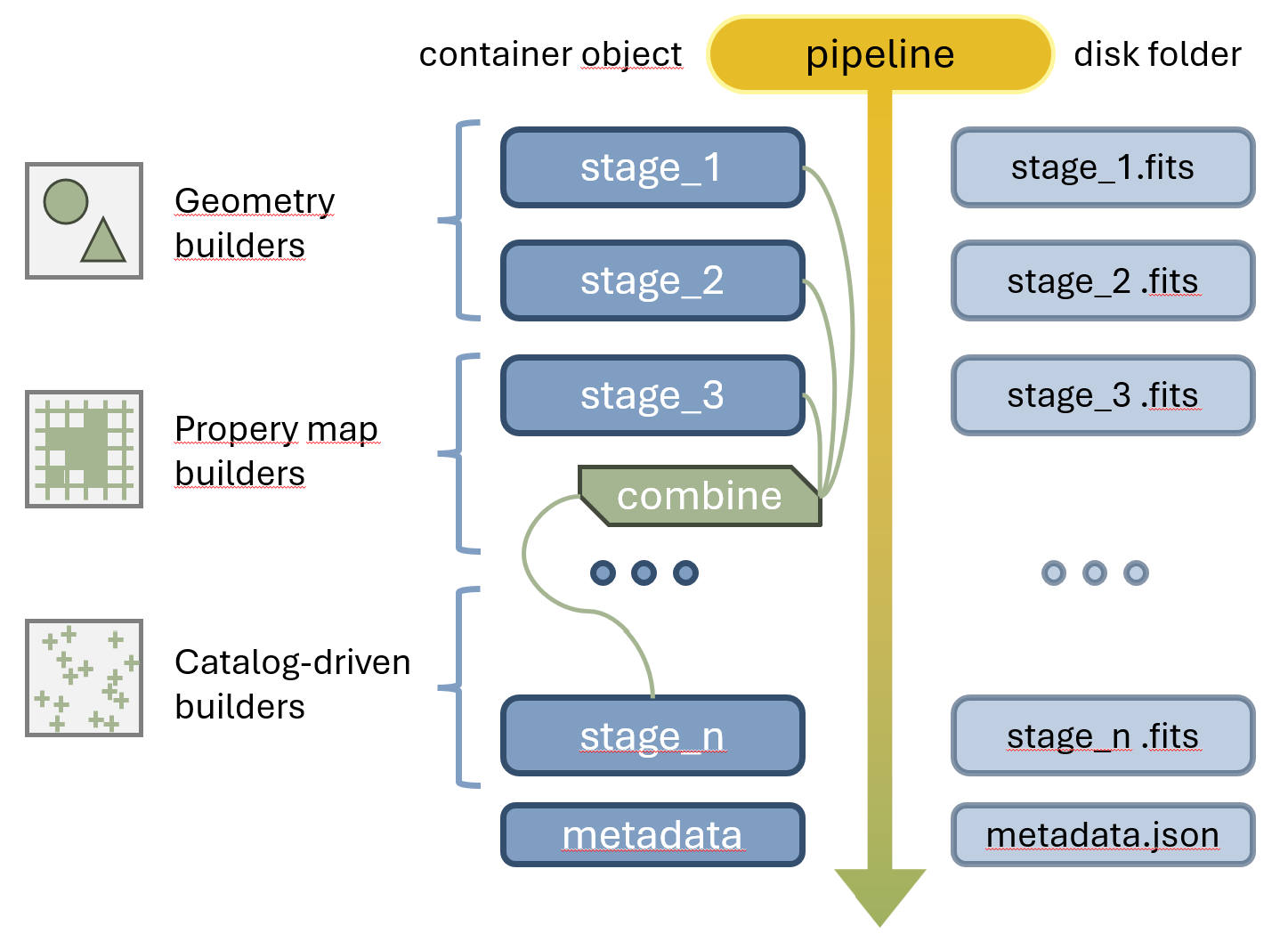}
    \caption{Schematic view of pipelines and stages in Skykatana.}
    \label{fig:schematic}
\end{figure}

Each Skykatana stage is internally a map whose values are boolean, but the code supports two storage modes implemented by HEALSparse. In standard mode, every sparse pixel is stored in an explicit boolean array. In bit-packed mode, up to eight pixel flags are encoded into a single byte, yielding an eightfold reduction in memory and I/O volume while preserving exact geometry. Bit packing does not affect logical operations, and most core methods transparently accept either format.

\subsubsection{Implementation and distribution}
Skykatana is implemented entirely in Python ($\geq~3.10$) and distributed as an installable package on PyPI under the name skykatana. It can be installed with a single command (\texttt{pip install skykatana}) and requires the HATS/LSDB libraries along with standard scientific dependencies (numpy, pandas, astropy, healsparse, and mocpy). The package is fully open source under the MIT license, hosted on GitHub, and accompanied by comprehensive online documentation and example notebooks. To ensure reproducibility, the exact version used in this work has been archived with a persistent DOI through Zenodo (DOI: \href{https://doi.org/10.5281/zenodo.18407348}{10.5281/zenodo.18407348}). The archive also includes a conda environment file and pinned dependency versions.

\subsection{Core algorithms and methods}
In the following subsections, we describe the various types of mask builders and highlight their most relevant characteristics. 

\subsubsection{Geometry-based builders}
These methods allow the creation of pixel sets from analytic shapes commonly used in astronomy, such as circles, ellipses, polygons, and zones (i.e., rectangular regions bounded by major circles along right ascension and minor circles along declination). Internally, the algorithms employ fast pixelization routines available in mocpy to generate its MOC representation, which is then flattened to the sparse order requested and converted to a boolean HEALSparse map. Example routines are \texttt{build\_zone\_mask()}, \texttt{build\_ellip\_mask()} and \texttt{build\_circ\_mask()}. As the latter method is meant to process hundreds of millions of stars, it includes chunked processing logic to pixelize and stream bursts of pixels directly into the output mask, therefore keeping the peak memory footprint nearly constant with sky area. This class of builders is typically used for user-defined masks or for extended sources whose angular shapes are well approximated by analytic forms.

\subsubsection{Property-map builders}
Methods such as \texttt{build\_prop\_mask()} can threshold multiple property maps (for instance, coadded depth, seeing, background variance, observing time) as those produced by data management teams of modern surveys, and return a combined map for the resulting valid/invalid pixels. This allows fast construction of quality masks from existing HEALSparse maps.

\subsubsection{Catalog-driven builders}
In some cases, pixelizing extensive surveys can be impractical due to the sheer size of the catalogs, or technical difficulties to estimate precise coverage regions. In some cases like Rubin the dataset will be too large to fit into memory without a significant computational infrastructure. To accommodate such cases, the code implements a method (\texttt{build\_foot\_mask()}) to directly pixelize a dense catalog of discrete sources. When a dataset has been partitioned with the HATS/LSDB framework it is already organized into roughly equal-size partitions (or HATS pixels) and materialized as a set of parquet files which can be loaded and processed in turn by one or more workers of a Dask cluster \citep{dask}. Skykatana wraps its pixelization logic as a function that is mapped to every partition, sent for execution across the cluster, and collects (reduces) the pixels at the end. In this way, an enormous task is divided into smaller pieces, each distributed across partitions, achieving out-of-core scalability with limited resources.

When pixelating a catalog of discrete sources, there is always a trade-off between contiguity and pixel size (i.e. order). In some cases, it might be preferable to apply post-processing operations to fix artificial small holes in the coverage or improve the matching around borders. Skykatana offers an option to search and perform a connectivity-based cleanup of single isolated empty (False) pixels within the main mask body. This is useful when pixelization leaves single-pixel artifacts in a contiguous area. In addition, the code can also apply a morphological erosion algorithm that trims one layer of pixels along the mask edge (optionally repeated several times), effectively shrinking the masked area by one or more HEALPix cells. This can mitigate narrow overlaps between adjacent exclusion zones or enlarge the jagged boundaries around empty (False) regions but at the same time increases the correspondence between pixelized areas and the real coverage of these regions.

\subsection{Star mask builder: scaling pixelization on demand}
\label{sec:starmaskbuilder}
We also include a detailed description of a special method to construct bright star masks. \texttt{build\_star\_mask\_online()} is designed to build an exclusion mask by querying an external stellar catalogs (e.g. Gaia in HATS/LSDB version) \emph{on demand} over an arbitrary sky region, then pixelizing circular exclusion zones around each star using a given radius relation (for instance, the empirical radius--magnitude relation from Sec.~\ref{sec:radius}). Compared to precomputed, monolithic stellar masks, this online approach offers several advantages: (i) users can tailor magnitude limits, radius prescriptions, and sky coverage without regenerating global products; (ii) throughput and memory are bounded by \emph{chunked} processing of the target region; (iii) avoids downloading and querying large catalogs locally, which can involve significant amount of time and resources; (iv) the method naturally supports distributed execution (e.g. within the Rubin Science Platform, see \citealt{LDM542-RSPDesign}, \citealt{LSE319-RSPVision}, \citealt{omullane2024}).


\subsubsection{Search-stage definition and MOC construction}
Users provide a \emph{search stage} that defines where stars should be queried and ultimately masked out. Most often, this is the survey footprint or a quality-cut mask. The stage is converted into its MOC representation at a user-specified maximum order $o_{\max}$, yielding a set of nested HEALPix cells $\mathcal{M}_{\mathrm{search}} = \{(o_i, p_i)\}$ whose union approximates the stage to subpixel precision.

\subsubsection{Breadth-first chunking of the search MOC}
To limit memory and I/O, $\mathcal{M}_{\mathrm{search}}$ is partitioned into work units by a breadth-first traversal of the MOC tree. The algorithm expands coarse cells only where necessary so that each chunk accumulates a nearly constant area or predicted source count. This breadth-first (BF) scheme produces compact, contiguous sky patches, large enough for efficient catalog queries, yet small enough to stream through memory safely. It also preserves coherence near boundaries, avoiding the fragmentation typical of depth-first partitions.

\subsubsection{Border enlargement}
Each chunk is padded \emph{before} catalog querying by expanding its angular
extent by $\Delta\theta$, typically a few arcminutes, chosen to exceed the
maximum expected radius $R_{\max}$ of the brightest stars. Padding ensures that stars whose masking discs overlap the chunk area but whose coordinates are strictly outside the chunk area are not missed. In practice, adding an order 8 pixel ($\sim$\ang{;;822} size) around each chunk is sufficient to include the effect of stars up to $\sim$\ang{;;411} radius.

\subsubsection{Milky Way masking during search}
Because stellar surface densities in the Galactic plane are extremely high, querying stellar catalogs over the full sky can become prohibitively time-consuming. Moreover, masking sources in these crowded regions is often scientifically unwarranted. To address this, a geometric Milky Way mask, denoted $\mathcal{M}_{\mathrm{MW}}$, can optionally be applied to exclude low Galactic latitudes—typically $|b| < 10^\circ$–$20^\circ$—as well as the Galactic bulge, modeled as an elliptical region. Both components have configurable parameters that allow users to adjust the extent of the excluded area. Operationally, the effective search region becomes $\mathcal{M}'_{\mathrm{search}} = \mathcal{M}_{\mathrm{search}}
 \setminus \mathcal{M}_{\mathrm{MW}}$ and only this reduced sky area is partitioned into BF chunks. This exclusion prevents unnecessary queries for hundreds of millions of stars that would otherwise dominate computation time and generate heavily perforated masks near the Galactic plane. The Milky Way mask is retained as an additional stage within the pipeline, allowing users to perform further operations if needed.

\subsection{Random point generation}
Skykatana generates random points uniformly over masks using a two-pass streaming algorithm that avoids materializing the full list of valid pixels. The mask is iterated in chunks grouped by coverage pixel. In a first pass, the number of valid sparse pixels per chunk is counted, and the requested total number of random points is distributed across chunks via a multinomial draw, proportional to the valid area. In a second pass, parent sparse pixels are sampled uniformly within each chunk and refined to a finer grid (by default \texttt{nside\_randoms=23}) by appending random sub-pixel bits, yielding uniform sampling within each sparse pixel. This design ensures bounded memory usage and scalability to high-resolution, highly perforated masks.

We quantify uniformity using a Kolmogorov--Smirnov (K--S) test at the native sparse resolution. Each random point is mapped to its sparse pixel index and converted to a continuous rank statistic over valid pixels. We choose a highly perforated mask (the join of LMC r--band masks including the halo component and faint Gaia stars, see Appendix A). Using \(10^7\) random points and 20 independent seeds, we obtain K--S statistics $D\sim2-4\times 10^{-4}$, consistent with the expected $\frac{1}{\sqrt{n}}$ behavior and showing no statistically significant deviation from uniformity.

As an independent test, we compute the angular two-point correlation function $w(\theta)$ using the Landy--Szalay estimator, with independent random catalogs ($N_D = 2\times10^6$, $N_R = 2\times10^7$). Over the fitted angular range from $1''$ to $1^\circ$, $w(\theta)$ is consistent with zero, with $\chi^2/\text{dof} = 0.16~(p = 0.99998)$ for the null hypothesis $w(\theta)=0$.

\subsection{Visualization of stage masks}
Skykatana provides several complementary tools to visualize, validate and interactively inspect individual mask stages. Because masks may contain up to billions of pixels, direct plotting of every pixel is rarely practical. Instead, Skykatana uses multiple lightweight representations that can be generated and displayed efficiently even for survey-scale data. 

\subsubsection{MOC and static quick-look views}
Each stage can be visualized through its MOC representation (using the \texttt{plot\_moc()} method). The code computes the minimal MOC footprint at the desired maximum order that is worth plotting, given the chosen viewport, and displays it as a set of filled HEALPix cells in a World Coordinate System (WCS) projection. Not less important, it also clips pixels to the current viewport, allowing fast rendering of large footprints. It also supports multiple overlaid stages, enabling direct comparison between. These plots provide a rapid diagnostic of the stage alignment, coverage, and relative contribution of each component to the final mask.

\subsubsection{Visualization by random points}
A complementary mask visualization method is through random point sampling. As said before, Skykatana can uniformly draw random coordinates within the valid (unmasked) region of a stage or of the final combined mask. These randoms are then plotted to verify uniformity and continuity, and to identify potential discontinuities or holes introduced by masking operations. Because only the random points are displayed rather than the underlying boolean map, this method allows visual inspection of very high-order masks with negligible memory cost and high performance.

\subsubsection{Interactive exploration}
Skykatana integrates seamlessly with the ipyaladin Jupyter widget, a Python interface to the Aladin Lite interactive sky atlas (\citealt{bochfernique2014}), enabling interactive visualization and inspection of mask stages directly within Jupyter notebooks. Stage masks can be automatically exported as MOCs and displayed in an ipyaladin viewer layer, enabling users to pan, zoom, and overlay catalogs or images directly from the notebook. This interface is particularly useful for examining the correctness of star-exclusion radii, verifying geometry boundaries, or cross-checking against sky images.

\subsubsection{Fractional area maps}
Beyond binary views, Skykatana can summarize a single boolean stage or the final combined mask into a fractional coverage map at a coarser HEALPix order, using highly efficient methods already implemented in HEALSparse. Concretely, a high-resolution sparse map is reduced to a lower order; and for each coarse coverage pixel, it computes the fraction of its sparse resolution children that are valid. The result is a floating point map that highlights perforation gradients due to stars, reveals partially filled tiles, and highlights crowded regions by dense objects such as globular clusters or MW satellite galaxies. By thresholding a fractional map at a given value, it is straightforward to create masks that tolerate up to a desired perforation level. We refer the reader to Sect. \ref{sec:examplerubin} for an analysis of the fractional area masked by Gaia stars within the Rubin footprint.

\subsection{Logical combination of stages}
Skykatana implements a special combination algorithm (\texttt{combine()}) that merges two or more stages into a single boolean mask at an output sparse order, even when stages in a pipeline are (possibly) defined at different coverage or sparse resolutions. To ensure consistency, the algorithm performs an on-the-fly reprojection of all inputs to the target sparse order. Such a reprojection is hierarchical and uses the native NESTED indexing of HEALPix, guaranteeing exact geometric containment and bit-identical results across different orders.

Once all stages share a common resolution, the code evaluates a logical expression tree that encodes the user-requested operation (e.g., union, intersection, and subtraction). The evaluation is executed in streaming mode, where pixels are read and combined in blocks rather than materialized as full arrays. In this way, memory usage scales with the number of active coverage rows rather than total sky pixels. This allows operations on maps containing billions of valid pixels with only a few gigabytes of RAM.

Suppose that the combined mask is requested at a lower order than the inputs. In that case, a hierarchical reduction is applied, marking a parent pixel as valid if any of its children are valid (for union), or only if all children are valid (for intersection). This preserves the logical semantics under down-sampling. Users can chain several logical combinations in a single call by providing an ordered list of stages to be intersected or added, and another list of stages that will be subtracted. The resulting map is returned as a new stage attached to the pipeline along with the corresponding metadata.

\subsection{IO and Interoperability}
Storing multiple large masks composed of billions of pixels with limited memory resources can be challenging, and simply saving a long list of (integer) pixel values is far from ideal. Skykatana stores a pipeline object in a single directory, one FITS table per stage, using a compact bit-packed boolean encoding designed for high-throughput streaming I/O and compact file size. As stages are discovered dynamically from the in-memory object, it is assured that they will be persisted to disk.

Instead of storing a dense boolean array per coverage pixel (wasteful when occupancy is sparse), SkyKatana writes only the minimal number of bytes needed to represent the highest child offset, with bits set for its active children. Then, in practice, each stage data is stored incrementally as a single-table FITS file with one row per coverage pixel that has at least one child, i.e. a sparse pixel set to True, and the pixel data itself of each coverage row is saved as a bit-packed (little-endian) byte array. Such a packing schema reduces the I/O volume and memory pressure while preserving exact geometry and compact file size. For reference, a single stage of 1.5 billion pixels can typically be written and read back in less than a minute, taking $\sim$600~MB of space on disk.

The Skykatana FITS format consists of a binary table with columns COVPIX (coverage pixel index), ENC (encoding flag), and PACKED (bit-packed byte array of child pixels). The table header records the sparse and coverage geometry and encoding details using the keywords listed in Table~\ref{tab_fitskeywords}. These keywords enable third-party readers to interpret the bit-packed geometry unambiguously using standard FITS tools. To facilitate interoperability, Skykatana supports exporting masks to other formats (e.g. MOC). Moreover, because in-memory stages are represented as HEALSparse maps, they can also be exported to the native HEALSparse FITS format when full compatibility with external HEALSparse workflows is required.

In addition to the FITS stage files, each pipeline directory contains a JSON metadata file that (i) records the parameters and methods used to generate each stage, including the number of valid pixels and total masked area, and (ii) stores the information needed to restore the pipeline state and reattach stages under their original names when reloaded.

\begin{table}[hb]
\centering
\begingroup
\begin{adjustbox}{max width=\linewidth}
\begin{tabular}{ll}
\hline
Keyword & Meaning \\
\hline
NSIDE\_COV & NSIDE of the coverage grid \\
NSIDE\_SPA & NSIDE of the sparse grid \\
NFINE & Children per coverage pixel: $(\mathrm{NSIDE\_SPA}/\mathrm{NSIDE\_COV})^2$ \\
DTYPE & Stored logical type (currently \texttt{'bool'}) \\
ENCOD & Encoding scheme (currently \texttt{'BITPACK'}) \\
BITORD & Bit order within bytes (currently \texttt{'L'} for little-endian) \\
\hline
\end{tabular}
\end{adjustbox}
\endgroup
\caption{Skykatana stage FITS header keywords.}
\label{tab_fitskeywords}
\end{table}

\section{Stellar magnitude-radius relations from Rubin DP1 data}
\label{sec:radius}

\begin{figure}[htbp]
    \centering
    \includegraphics[width=0.48\textwidth]{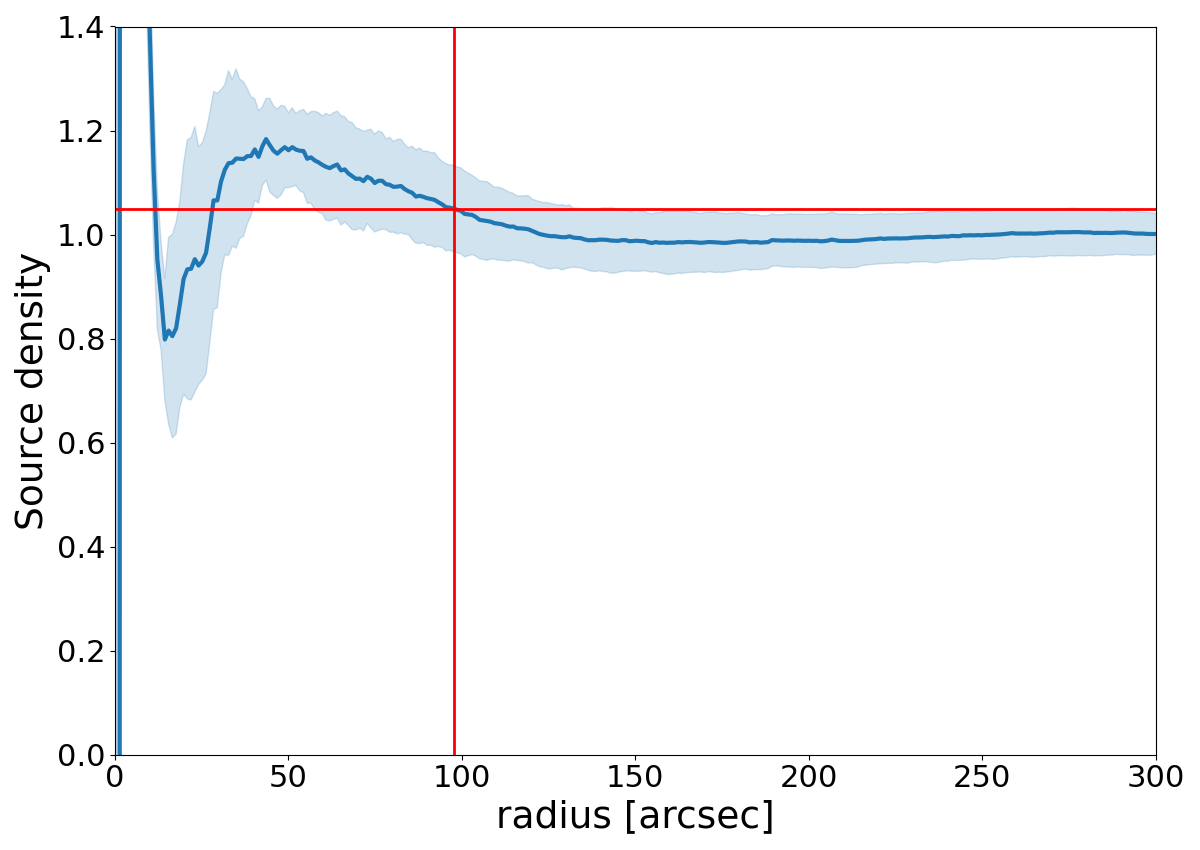}\\[0em] 
    \includegraphics[width=0.48\textwidth]{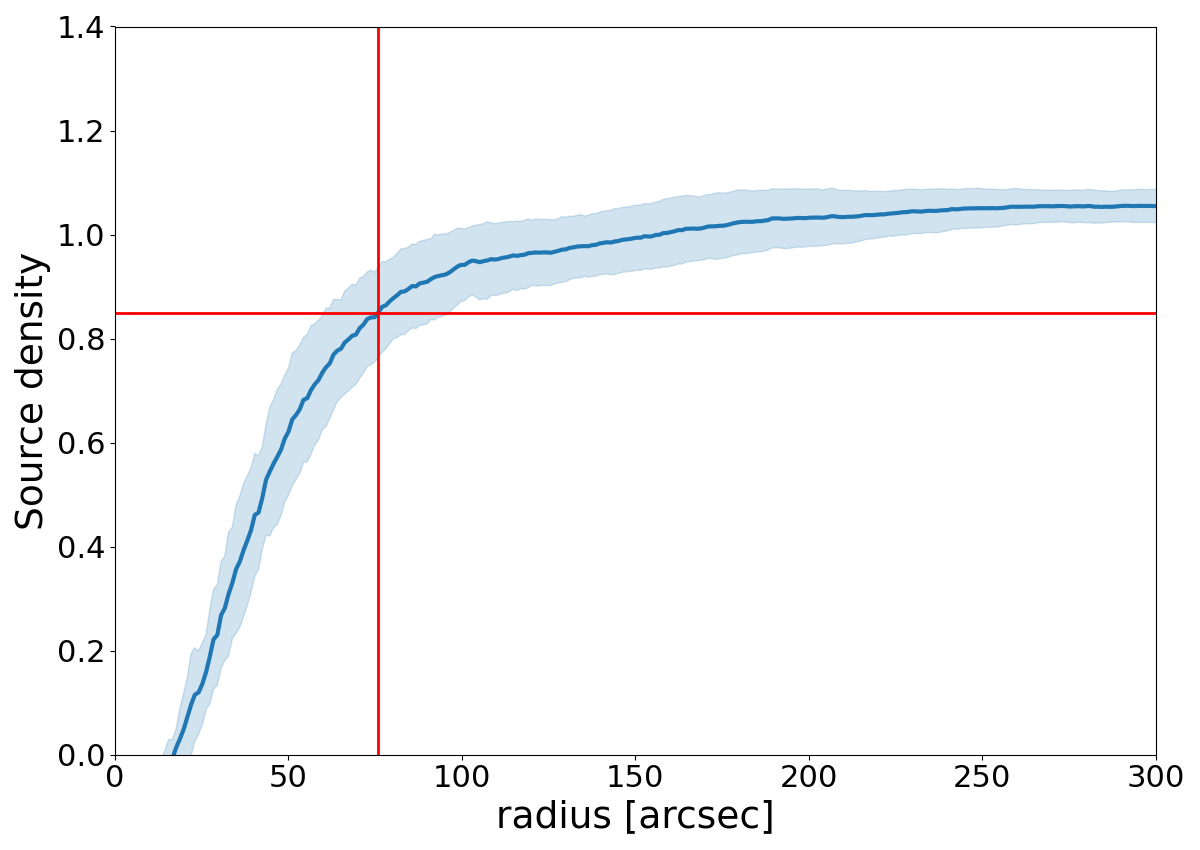}\\[0em]
    \caption{Mean radial density profiles of DP1 sources around Gaia stars in the same magnitude bin ($10.3 < G_{\mathrm{Gaia}} < 11.6$). The top panel shows the profile obtained when all detections are included, displaying the characteristic features of bright-star contamination: a spurious central overdensity from core fragmentation, a depletion region caused by incompleteness within the stellar halo, and a secondary excess associated with the fragmented outer halo.The bottom panel shows the corresponding profile after removing halo-related artifacts using \texttt{blendedness\_flag == False}, yielding a smooth, monotonic increase toward the field density. Shaded regions in both panels represent the $1\,\sigma$ dispersion across stars in the bin.}
    \label{fig:density_profile}
\end{figure}

\subsection{Data pre-processing}
To calibrate the stellar exclusion radii, we used Rubin DP1 photometry (\citealt{RTN095-DP1}) produced by LSST pipeline \citep{10.1093/pasj/psx080} within the Extended Chandra Deep Field South (ECDFS).
Among the DP1 fields, ECDFS was selected because it is the deepest LSST region currently available and targets a relatively clean extragalactic area, making it representative of low-crowding conditions relevant for cosmological analyses. In such regions, source detection and completeness are less affected by overlapping stellar halos, enabling a more robust characterization of radial density profiles around bright stars. By contrast, in highly crowded fields we expect these profiles to be noisier due to the effect of overlapping halos and source confusion, which would complicate the determination of well-defined exclusion radii.

Our first goal was to construct a background distribution of sources with approximately uniform spatial completeness. For this, we made use of the \texttt{deepCoadd\_psf\_maglim\_consolidated\_map\_weighted\_mean} property maps \citep{2025arXiv250105739B} available in the Rubin Science Platform\footnote{\href{https://dp1.lsst.io/tutorials/notebook/203/notebook-203-1.html}{dp1.lsst.io/tutorials/notebook/203/notebook-203-1.html}}, which store the depth at HEALPix resolution as HEALSparse maps. These property maps allow us to determine the maximum depth reached within the ECDFS for each band, and we adopted this value as the faintest usable magnitude for the DP1 sources. These limits correspond to 25.39, 27.38, 27.12, 26.53, 25.67, and 23.77 in the \textit{u, g, r, i, z, y} bands, respectively. Next, to remove shallow border regions and ensure spatial uniformity, we used the \texttt{build\_prop\_mask} function of Skykatana to generate a boolean mask selecting only those pixels where the local depth exceeds a band-dependent threshold. The thresholds were 24.5, 26.5, 26.0, 25.5,  25.0, and 23.0 in \textit{u}, \textit{g}, \textit{r}, \textit{i}, \textit{z}, and  \textit{y}, respectively. Only DP1 sources within these masks were included in the analysis.

Gaia stars were selected in the magnitude range $9 \le G_{\mathrm{Gaia}} \le 21$. Stars brighter than $G\simeq9$ were excluded because they are too few  within ECDFS to define statistically robust radii. Only Gaia stars located within the footprint defined by the corresponding depth cut and at least $400''$ away from the field boundaries were considered to avoid edge effects that could bias the density profiles and fitted radii. For each star, we computed the radial profile of DP1 source counts and normalized it by the mean density of the selected background region. A mean profile was then obtained for each magnitude bin by averaging the individual profiles, while the standard deviation across stars in the bin served as the uncertainty estimate.

A well-known complication in wide-field survey pipelines is that very bright stars can produce numerous artificial detections due to over-deblending of their light profiles, both in the saturated core and in the extended PSF wings. These spurious sources arise when source detection and deblending algorithms interpret complex light structures as multiple discrete objects, leading to fragmented catalogs around bright stars (\citealt{melchior2018}, \citealt{Mullaney2021}), producing  over-densities in the vicinity of bright stars. The LSST \texttt{blendedness\_flag} is raised when the pipeline is unable to compute a reliable blendedness metric for a given detection. This typically occurs under non-ideal photometric or geometric conditions such as saturated stellar cores, extremely steep flux gradients, irregular footprints, or severe contamination where the algorithm cannot robustly determine the relation between the flux of the source and that of its inferred parent source. The \texttt{blendedness\_flag} is provided in the LSST Object table. In practice, this flag correlates strongly with spurious deblender fragments around bright stars, and selecting only objects with \texttt{blendedness\_flag == False} effectively removes the majority of these artifacts. Although this cut eliminates some real faint galaxies projected over bright halos, its impact on the averaged profiles is modest. For this reason, we carried out \emph{two parallel analyses}: one including all DP1 sources and thus preserving the halo-induced overdensities, and one excluding halo artifacts via \texttt{blendedness\_flag == False}. Figure~\ref{fig:density_profile} shows two representative mean profiles. When all sources are included, the profile exhibits a spurious central excess, followed by a depletion where the star suppresses faint detections, and eventually a secondary excess associated with halo fragmentation. When halo artifacts are removed, the profile increases smoothly and monotonically toward unity. These qualitative differences motivate the distinct definitions of stellar radius described below.

\begin{figure}[htbp]
    \centering
    \includegraphics[width=1\linewidth]{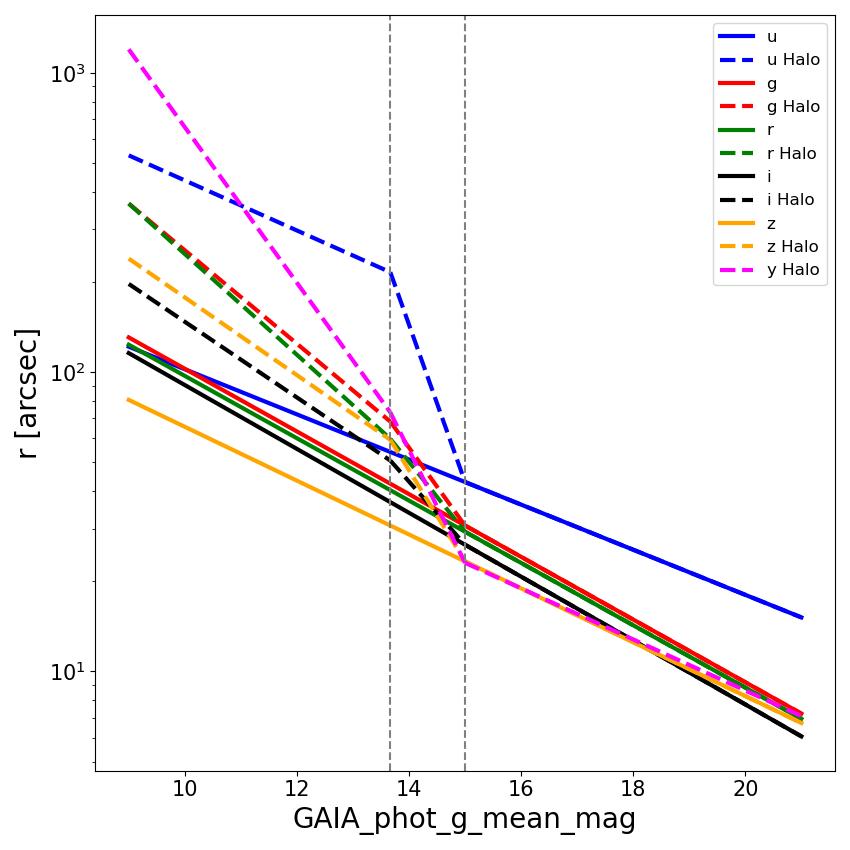}
    \caption{Fitted radius–magnitude relations for all LSST bands in the halo and no-halo cases. For the halo case, three power-law segments are fitted per band, with boundaries at $G_{\mathrm{Gaia}} = 13.6$ and $G_{\mathrm{Gaia}} = 15.0$, capturing the transition from strong halo contamination to the weak-halo regime. The no-halo case is well described by a single power law per band. Best-fit parameters are listed in Tables~\ref{tab:halo_band_parameters} and \ref{tab:no_halo_and_parameters}.}
    \label{fig:radius_vs_mag}
\end{figure}

\subsection{Model and Fitting}
\label{sec:Model_and_Fitting}

The different behavior of the profiles with and without halo contamination necessitates two distinct definitions of the stellar exclusion radius. When the halo is included, the relevant boundary is the outer edge of the halo-induced overdensity, and we define the radius as the point where the mean profile reaches 1.05, i.e.\ 5\% above the field density. When the halo is removed, the radius corresponds instead to the onset of  completeness recovery, defined as the radius where the profile reaches 0.85.

Uncertainties were obtained by intersecting the upper and lower $1\sigma$
envelopes of the mean profile with the appropriate threshold, as shown in
Fig.~\ref{fig:density_profile}. This procedure captures the variability among 
stars within each magnitude bin.

\begin{figure}[htbp]
    \centering
    \includegraphics[width=1\linewidth]{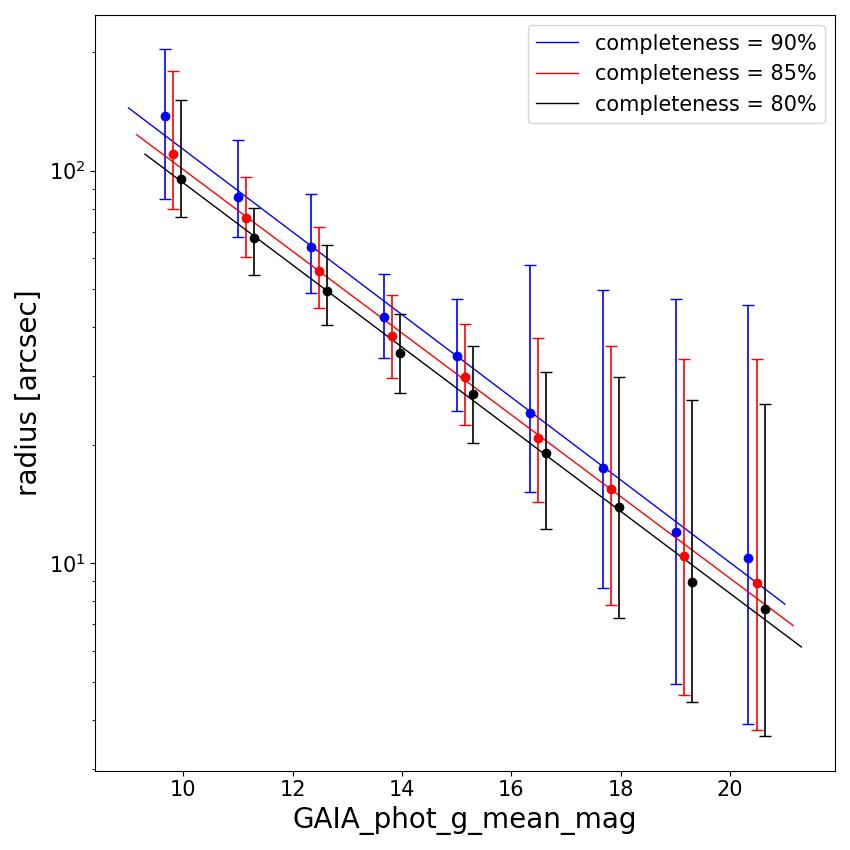}
    \caption{Best-fitting radius–magnitude relations in the $r$ band for three completeness thresholds  (90\%, 85\%, and 80\%), computed after removing halo artifacts. The corresponding fitted parameters are listed in Table~\ref{tab:r_completeness}}
    \label{fig:radius_vs_mag_completeness}
\end{figure}

\begin{figure}[htbp]
    \centering
    \includegraphics[width=1\linewidth]{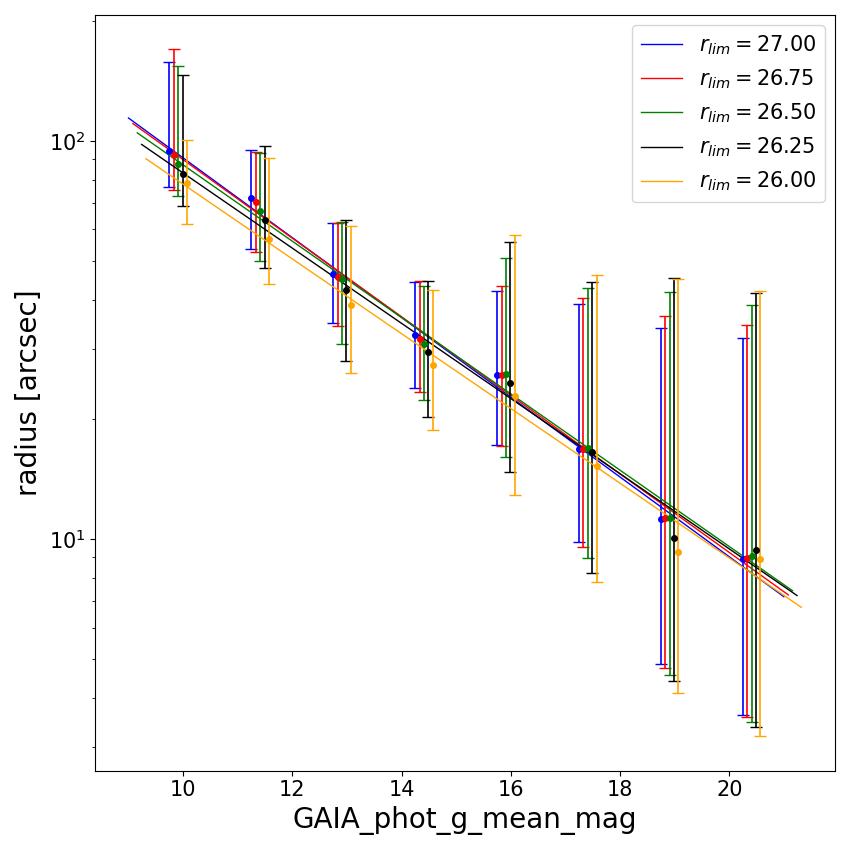}
    \caption{Radius–magnitude relations in the $r$ band for different limiting magnitudes ($r_{\mathrm{lim}} = 27.00, 26.75, 26.50, 26.25, 26.00$) used when computing background densities, all adopting the 85\% completeness threshold. The fitted parameters for each limit are given in Table~\ref{tab:limit_mag_parameters}}
    \label{fig:limit_mag_parameters}
\end{figure}

When halos are included, the radius--magnitude relation shows three distinct regimes. For bright stars ($G\lesssim13.6$) the radii decrease rapidly with magnitude, reflecting the shrinking extent of the saturated halo. For fainter stars ($G\gtrsim15$) the halo is negligible and the radii follow a shallower trend. Between these limits, the relation transitions smoothly. For each band, we therefore fitted three consecutive power laws of the form
\begin{equation}
R(G_{\mathrm{Gaia}}) = A_i\exp(B_i G_{\mathrm{Gaia}})
\label{eq:1}
\end{equation}

one in each magnitude interval, enforcing the continuity at $G_{\mathrm{Gaia}}=13.6$ and $G_{\mathrm{Gaia}}=15.0$. In this empirical relation, $A_i$ serves as the normalization constant defining the scale of the exclusion zone, while $B_i$ determines the rate at which the radius shrinks with increasing stellar magnitude, physically capturing the steepness of the PSF wings and the resulting halo contamination profile in each filter. To further isolate the halo signal, these fits were performed using only DP1 sources brighter than the conservative magnitude limits: 23.75, 25.00, 25.00, 25.00, 24.20, and 23.00 in \textit{u, g, r, i, z, y}, respectively. When the halo is removed, the radius--magnitude relation becomes smooth across the full magnitude range and is well described by a single power law per band. Tables~\ref{tab:halo_band_parameters} and \ref{tab:no_halo_and_parameters} summarize the fitted parameters for the halo and no-halo cases, respectively, and Fig.~\ref{fig:radius_vs_mag} shows the corresponding fitted relations. In the \textit{y} band, residual halo-induced overdensities persist even after quality cuts, preventing a reliable fit.

\begin{table}[htbp]
\centering
\begin{adjustbox}{max width=\linewidth}
\begin{tabular}{c r r r r r r} 
 \hline
 Band & $A_1$ & $B_1$ & $A_2$ & $B_2$ & $A_3$ & $B_3$ \\ 
 \hline
 u & 2997.055 & -0.192 & 3297263555.368 & -1.210 & 579.537 & -0.173 \\
 g & 9283.754 & -0.359 & 251355.534 & -0.600 & 1139.957 & -0.241 \\
 r & 11960.396 & -0.387 & 91530.480 & -0.536 & 1068.425 & -0.239 \\
 i & 2706.861 & -0.291 & 38649.389 & -0.485 & 1056.128 & -0.245 \\
 z & 6174.420 & -0.437 & 6174.420 & -0.437 & 6174.420 & -0.437 \\
 y & 261334.672 & -0.598 & 261334.672 & -0.867 & 439.014 & -0.196 \\
 \hline
\end{tabular}
\end{adjustbox}
\caption{Fitted parameters of the radius–magnitude relations for the halo case in all LSST bands, with transitions at $G_{\mathrm{Gaia}} = 13.6$ and $G_{\mathrm{Gaia}} = 15.0$. The corresponding fitted relations are shown in the Fig.~\ref{fig:radius_vs_mag}.
}
\label{tab:halo_band_parameters}
\end{table}

\begin{table}[htbp]
\centering
\begin{adjustbox}{max width=\linewidth}
\begin{tabular}{c r r} 
 \hline
 Band & $A$ & $B$ \\ 
 \hline
 u & 579.537 & -0.173 \\
 g & 1143.921 & -0.241 \\
 r & 1068.425 & -0.239 \\
 i & 1056.128 & -0.245 \\
 z & 520.751 & -0.207 \\
 \hline
\end{tabular}
\end{adjustbox}
\caption{Best-fit parameters of the radius–magnitude relations for the no-halo case in all LSST bands except $y$, after removing halo-related artifacts using  \texttt{blendedness\_flag == False.}
}
\label{tab:no_halo_and_parameters}
\end{table}

We performed additional fits in the \textit{r} band, always without halo. First, we varied the completeness threshold used to define the radius (90\%, 85\%, 80\%). Lower completeness levels yield systematically smaller radii, with reductions of $\sim$5\% between the 90\% and 80\% criteria. The parameters appear in Table~\ref{tab:r_completeness}, and the curves in Fig.~\ref{fig:radius_vs_mag_completeness}. Second, we varied the upper magnitude limit used for DP1 sources ($r_{\mathrm{lim}}=27.00, 26.75, 26.50, 26.25, 26.00$). These fits appear in Table~\ref{tab:limit_mag_parameters} and Fig.~\ref{fig:limit_mag_parameters}. A clear trend is observed: as the limiting magnitude becomes brighter (i.e., when only relatively bright sources are included), the derived exclusion radii decrease, especially for the brightest Gaia stars. This effect arises because faint sources, near the detection threshold, are the first to reappear in the outer regions of the stellar halo. When these faint detections are excluded by a brighter magnitude cut, the radial density profile around bright stars becomes steeper: the local density remains depressed over a smaller area before returning to the field level. Consequently, the radius at which the normalized density profile reaches the 85\% criterion is smaller.

\begin{table}[htbp]
\centering
\begin{adjustbox}{max width=\linewidth}
\begin{tabular}{c r r} 
 \hline
 completeness & $A$ & $B$ \\ 
 \hline
 90\%  & 1280.150 & -0.242 \\ 
 85\%  & 1068.425 & -0.239 \\ 
 80\%  & 962.633 & -0.240\\  
 \hline
\end{tabular}
\end{adjustbox}
\caption{Best-fit parameters of the radius–magnitude relation in the $r$ band for three completeness thresholds (90\%, 85\%, and 80\%), using the no-halo density profiles (\texttt{blendedness\_flag == False}). The fitted curves are shown in Fig.~\ref{fig:radius_vs_mag_completeness}
}
\label{tab:r_completeness}
\end{table}

\begin{table}[htbp]
\centering
\begin{adjustbox}{max width=\linewidth}
\begin{tabular}{c r r} 
 \hline
 $r_{\mathrm{lim}}$ & $A$ & $B$ \\ 
 \hline
 27.00 & 911.438 & -0.230 \\ 
 26.75 & 855.403 & -0.227 \\ 
 26.50 & 764.632 & -0.220 \\  
 26.25 & 694.266 & -0.217 \\  
 26.00 & 631.991 & -0.216 \\  
 \hline
\end{tabular}
\end{adjustbox}
\caption{Fitted parameters of the radius–magnitude relation in the $r$ band for different limiting magnitudes used in the background-source selection ($r_{\mathrm{lim}} = 27.00, 26.75, 26.50, 26.25, 26.00$), all adopting the 85\% completeness definition and using no-halo density profiles. See Fig.~\ref{fig:limit_mag_parameters} for the corresponding fitted relations.
}
\label{tab:limit_mag_parameters}
\end{table}

\subsection{Results and Validation}
When halos are included, the \textit{u} and \textit{y} bands yield particularly large radii, sometimes exceeding several arcminutes for bright stars. This is expected given their shallow depth and the prominence of scattered-light structures in these filters. When halo artifacts are excluded, the relations become considerably more stable and physically plausible, though the \textit{y} band still shows residual contamination that prevents a reliable fit. Across the remaining bands, the faint-end slopes are broadly similar, following a slowly varying exponential decline, while the bright-end behavior varies significantly from band to band due to wavelength-dependent PSF structure and depth variations.

\begin{figure}[htbp]
    \centering
    \includegraphics[width=1\linewidth]{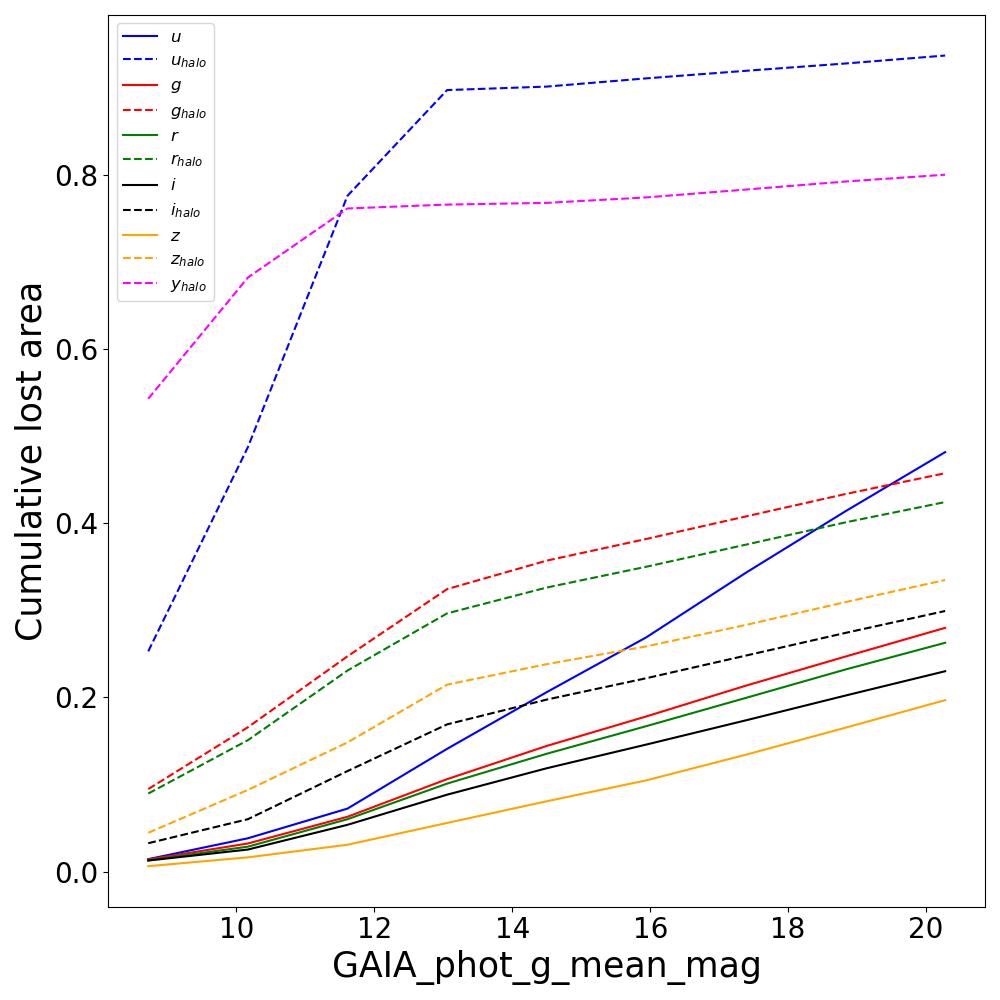}
    \caption{Cumulative masked-area fraction in the ECDFS as progressively fainter stars are included, using the halo (dashed) and no-halo (solid) radius–magnitude relations for each LSST band. Curves correspond to the 85\% completeness criterion. }
    \label{fig:Cumulative_lost_area_band}
\end{figure}

Using these relations, we computed the cumulative masked-area fraction in the ECDFS as progressively fainter stars are included. The resulting curves for the 85\% completeness case are shown in Fig.~\ref{fig:Cumulative_lost_area_band}. When halo contamination is included, the \textit{u} and \textit{y} bands yield significantly large masked fractions. With halo artifacts removed, the masked fractions are considerably reduced and more consistent across bands, although the \textit{u} band still shows higher losses due to its limited depth.

\begin{figure}[htbp]
    \centering
    \includegraphics[width=1\linewidth]{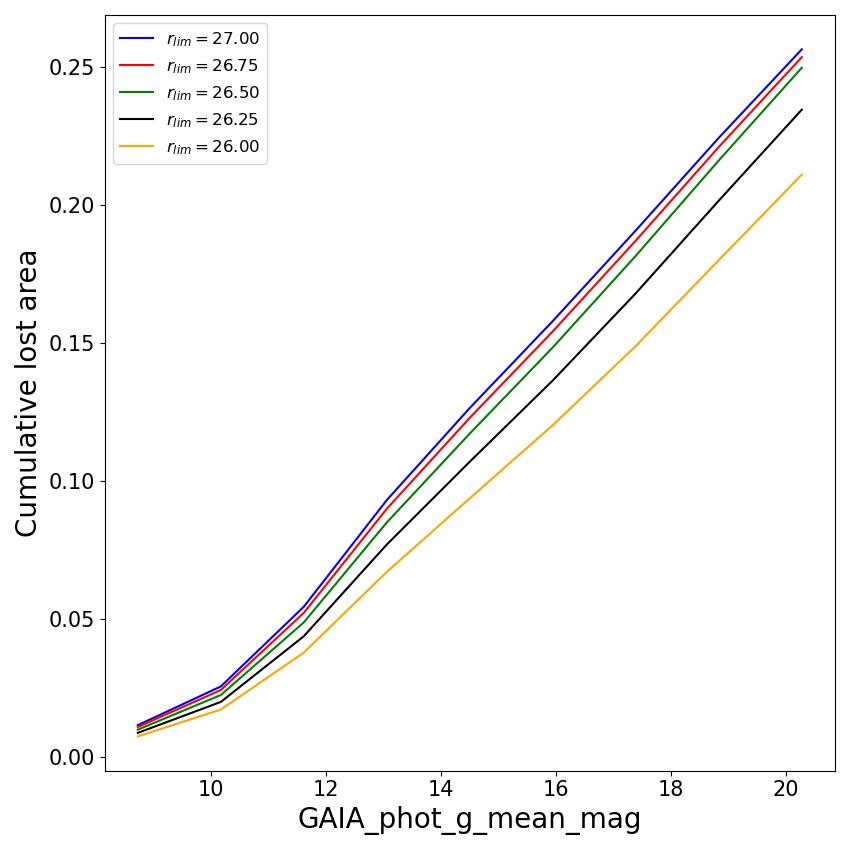}
    \caption{Cumulative masked-area fraction in the ECDFS for the $r$ band when adopting radius–magnitude relations derived using different limiting magnitudes for the background-source selection ($r_{\mathrm{lim}} = 27.00, 26.75, 26.50, 26.25, 26.00$).}
    \label{fig:Cumulative_lost_area_magcut}
\end{figure}

We also evaluated how the upper magnitude limit applied to DP1 sources affects the inferred masked area. The results, shown in  Fig.~\ref{fig:Cumulative_lost_area_magcut}, reveal that shallower magnitude limits lead to smaller masked fractions. This reflects the fact that faint detections near the edge of bright halos are excluded when imposing shallower limits, causing the inferred completeness recovery to occur at smaller radii.

To further quantify how the adopted magnitude limit in the density estimation affects the global masking efficiency, we repeated the area loss calculation using the radius–magnitude relations derived from the depth–cut analysis presented in Section~\ref{sec:Model_and_Fitting} (Fig.~\ref{fig:limit_mag_parameters} and Table~\ref{tab:limit_mag_parameters}). In this case, the cumulative masked fraction was recomputed for each set of radii corresponding to limiting magnitudes $r_{\mathrm{lim}} = 27.0, 26.75, 26.50, 26.25,$ and $26.0$, and the results are shown in Figure~\ref{fig:Cumulative_lost_area_magcut}. The figure reveals that shallower magnitude cuts, which consider only brighter background sources, lead to systematically smaller masked areas, most noticeably for the brightest stars. This trend is consistent with the behavior seen in the radius–magnitude relations of Fig.~\ref{fig:limit_mag_parameters}: when faint detections near the detection limit are excluded, the completeness around bright stars appears to recover at smaller radii, resulting in a reduced integrated masked area. For fainter stars, the effect is negligible, as their surroundings are only weakly affected by the choice of magnitude threshold. The dependence of the lost area on $r_{\mathrm{lim}}$ quantifies how sensitive bright–star masking is to the adopted catalog depth and detection completeness. For Rubin DP1, the relative variation in total masked area is of order $\lesssim15$--20\% between the different magntiude cuts analyzed here.

Overall, these empirical magnitude--radius relations, when coupled with \texttt{Skykatana} masking functionality, provide a reliable prescription for generating bright-star masks tailored to specific completeness targets and photometric bands, while highlighting the need for future calibrations based on wider survey areas that include a statistically significant sample of very bright stars.

Using Gaia and Tycho-2 stars over the full HSC footprint, \cite{coupon2018} derived empirical radius--magnitude relations based on the depletion of source density around saturated stars. Their adopted functional form consists of two exponential branches, which correspond to exclusion radii of $\sim50$--$60''$ for $G_{\mathrm{Gaia}}\simeq10$ and $\sim5''$ for $G_{\mathrm{Gaia}}\simeq17$. . Our ECDFS relations show broadly consistent trends: the exponential slopes at the faint end are comparable to those of HSC within uncertainties, while the bright-end radii are slightly smaller, reflecting the fact that the ECDFS data are derived from a more limited one-square-degree field. Overall, both studies demonstrate that empirical bright-star masking based on density profiles yield robust, survey-specific prescriptions that balance completeness and area loss.


We emphasize that the bright-star radius–magnitude relations derived in this section are based on a single $\sim1 \mathrm{deg}^2$ DP1 field (ECDFS), which provides limited statistics at the bright end and does not allow for a robust calibration in the $y$ band. Consequently, these relations should be interpreted as field-specific characterizations rather than universal prescriptions. Variations in survey depth, observing conditions, stellar density, and the effective area of the available fields are expected to introduce field-to-field differences. In particular, future analyses based on substantially larger sky areas will sample a broader population of bright stars and are therefore likely to yield different and more robust estimates of the characteristic exclusion radii.

\section{Application I: HSC–WISE composite mask}
For the first application of Skykatana, we will create a joint HSC-WISE mask to demonstrate its basic usage. For convenience, we will restrict ourselves to a small dataset of $\sim$8 million HSC galaxies (available \href{https://drive.google.com/file/d/1Fft9E9uD1eXs-8Dxb8bp5ou1bEtCTkgr/view?usp=sharing}{here}, including the necessary auxiliary files). This section follows the example notebook distributed in the online repository (available \href{https://github.com/samotracio/skykatana/blob/main/notebooks/quick_example_hsc.ipynb}{here}).

\subsection{Initial setup and source input}
We will first instantiate a pipeline object (\texttt{mkp}) and add a footprint stage for HSC, i.e. a stage defined by the pixelization of the discrete sources ingested as a HATS catalog (\texttt{HATSCAT}). We set an output order of 15 (6.4'' pixel size), but pixelize objects at order 13, which is better suited for a sample of this surface density.

\begin{lstlisting}[language=Python, caption={Initialization of the pipeline and pixelization of HSC sources}]
mkp = SkyMaskPipe(order_out=15)
mkp.build_foot_mask(sources=lsdb.open_catalog(HATSCAT), order_sparse=13)
\end{lstlisting}

\subsection{Adding a patch mask}
\begin{figure}[htbp]
    \centering
    \includegraphics[width=1\linewidth]{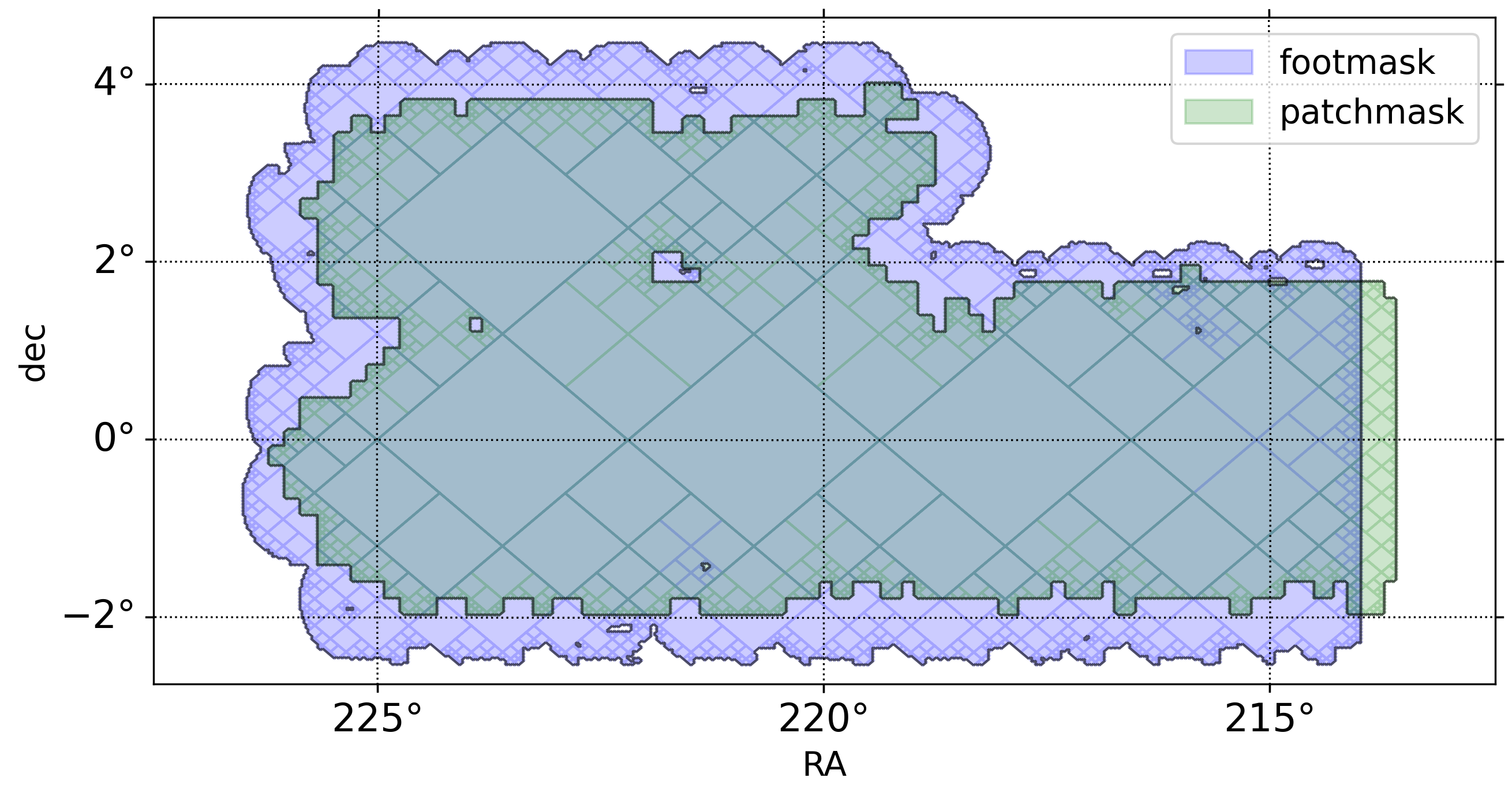}
    \caption{HSC-SSP footprint source mask and patch mask, for patches of depth $r_{psf}>25.9$ mag.}
    \label{fig:hsc_footpatch}
\end{figure}

The HSC-SSP survey is divided into patches (\texttt{PATCH\_FILE}) which are the fundamental areal units that result from diving each tract into smaller rectangular equi-area regions of $\sim$ 12' side. Several QA measurements (\texttt{QA\_FILE}) are available per patch. Below we create a patchmask stage after filtering out patches below a given PSF limiting magnitude in gri bands. Its MOC representations are shown in Figure \ref{fig:hsc_footpatch}.

\begin{lstlisting}[language=Python, caption={Adding a mask for HSC patches}]
filt = "(gmag_psf_depth>26.2) and (rmag_psf_depth>25.9) and (imag_psf_depth>25.7)"
mkp.build_patch_mask(patchfile=PATCH_FILE, qafile=QA_FILE, filt=filt, order_sparse=13)
\end{lstlisting}

\subsection{Masking bright stars}
The effect of bright stars in HSC can be decomposed into circular shapes due to the bright halo of stars (\texttt{STARS\_REGIONS}), and into rectangular boxes due to saturated bleed trails along the camera CCDs (\texttt{BOX\_STARS\_REGIONS}, see \citealt{coupon2018} for their definition).

\begin{lstlisting}[language=Python, caption={Adding masks for halos of bright stars and CCD bleeding.}]
mkp.build_circ_mask(data=STARS_REGIONS,fmt='parquet')
mkp.build_box_mask(data=BOX_STARS_REGIONS,fmt='parquet')
\end{lstlisting}

\subsection{Combine stages}
\begin{figure}[htbp]
    \centering
    \includegraphics[width=1\linewidth]{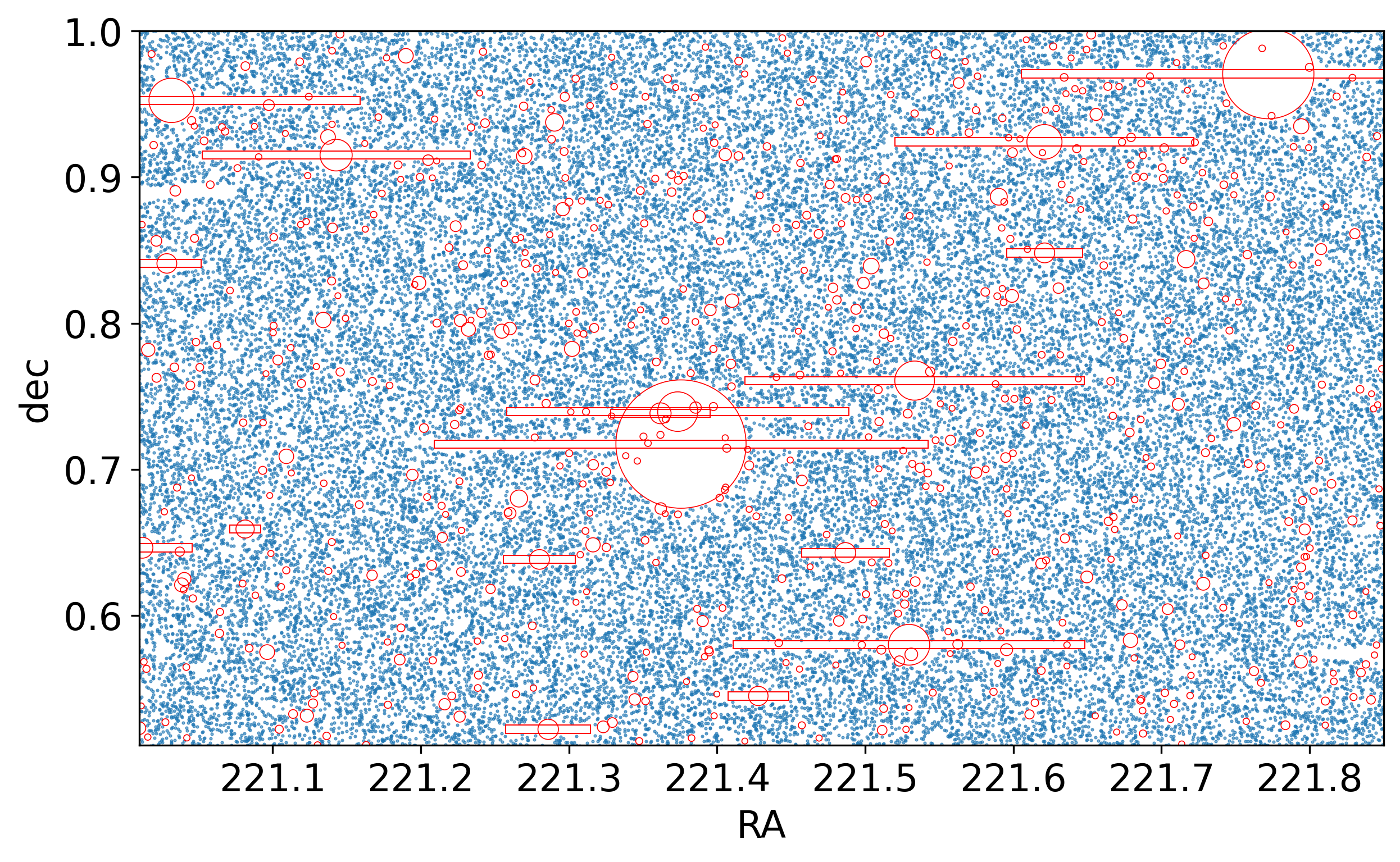}
    \caption{Visualization of the combined HSC mask by generating randoms sources within. The geometrical shapes excised by stars ares overlaid, confirming they are indeed excluded.}
    \label{fig:hsc_ranwbox}
\end{figure}

So far we have added various stages that are created with their default names and orders (all are customizable). Now we combine them by intersecting the HSC source-based footprint stage with the patch mask (specified as 'positive' stages); and subtracting the circles and boxes due to stars (specified as 'negative' stages). If maps have different geometry, such as sparse or coverage order, they will be reprojected on the fly to allow combination. In Figure \ref{fig:hsc_ranwbox} we visualize the combined mask by plotting random sources and overlaying the excised geometrical figures. As expected, no random points lie within the masked area.

\begin{lstlisting}[language=Python, caption={Combining stages.}]
mkp.combine(positive=[('footmask','patchmask')], negative=['circmask','boxmask'], verbose=True)
\end{lstlisting}

\subsection{Adding a stage for WISE}
\begin{figure}[htbp]
    \centering
    \includegraphics[width=1\linewidth]{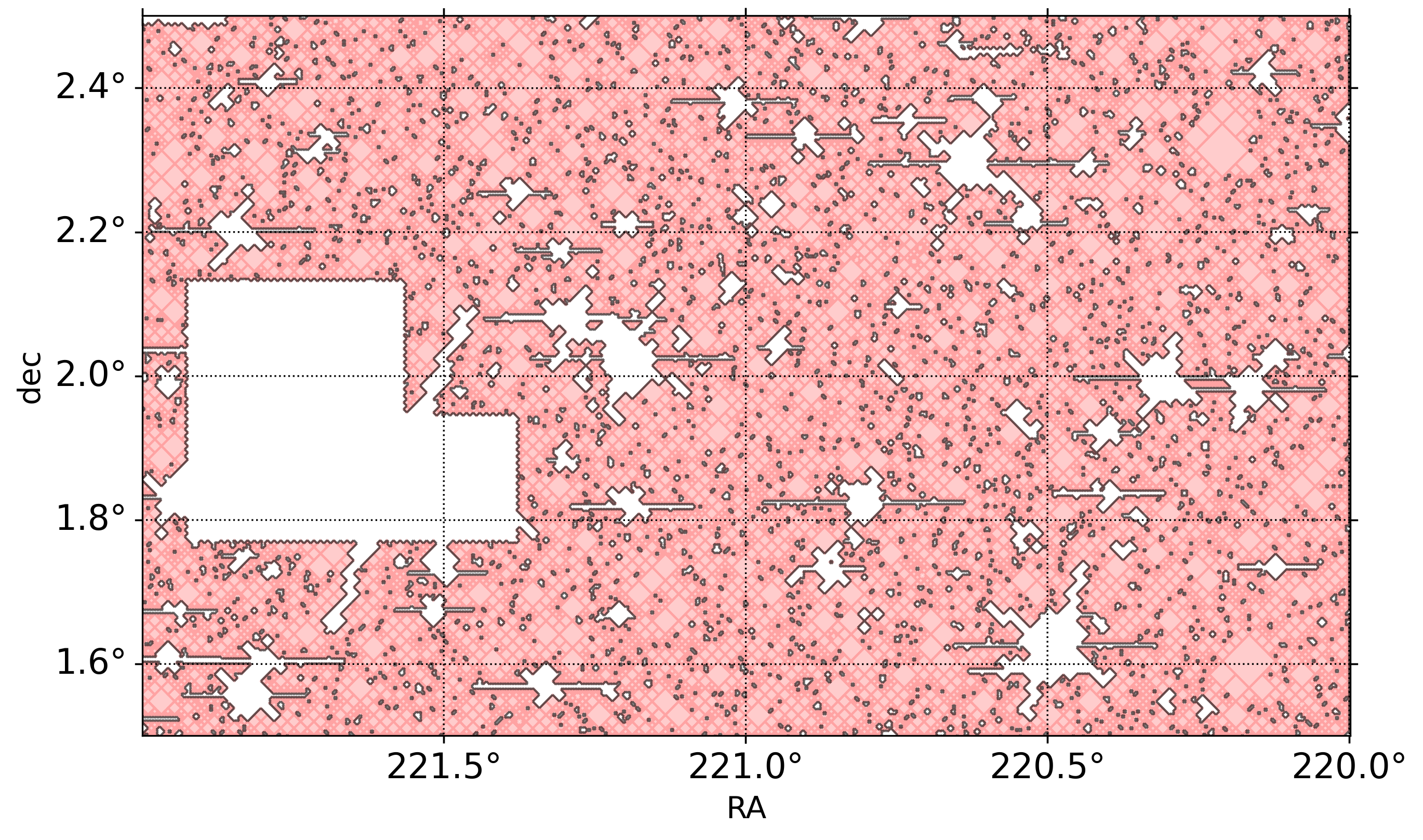}
    \caption{MOC visualization of the combined HSC+WISE mask. The (larger) diffraction spikes around stars traced in WISE are clearly visible. The large blank area follows from HSC patches discarded due to the presence of a bright star whose spikes are delineated in the WISE mask.}
    \label{fig:hsc_pluswise}
\end{figure}

To create a mask for WISE artifacts (difraction spikes, star halos, ghosts, etc.) we pixelize a catalog of objects severely affected by them. These were queried from the Reject table of the AllWISE database\footnote{\href{https://irsa.ipac.caltech.edu/Missions/wise.html}{irsa.ipac.caltech.edu/Missions/wise.html}}, by selecting objects either spurious or contaminated by those artifacts in W1, W2 or W3 bands. It is entirely possible to create a mask for moon trails (i.e. scattered light due to off-field lunar illumination) with the same procedure, but omit the steps for the sake of brevity. Below, we pixelize the artifacts (\texttt{WISE\_FILE}) and regenerate the combined mask. The resulting MOC is visualized in Figure \ref{fig:hsc_pluswise} and Table \ref{params_app1} lists the key configuration parameters adopted. 

\begin{lstlisting}[language=Python, caption={Combining stages.}]
mkp.build_foot_mask(pd.read_parquet(WISE_FILE), order_sparse=12, output_stage='wisemask')
\end{lstlisting}

\begin{table}[hb]
\begingroup
\begin{adjustbox}{max width=\linewidth}
\begin{tabular}{lc}
\hline
Parameter & Value \\
\hline
HEALPix sparse order (footprint) & 13 \\
HEALPix sparse order (output) & 15 \\
Coverage order & 4 \\
Bit-packed storage & Enabled \\
Logical combination & Footprint $\cap$ Patch $-$ Stars $-$ WISE artifacts \\
\hline
\end{tabular}
\end{adjustbox}
\endgroup
\caption{Key configuration parameters for the HSC--WISE demonstration (Application I).}
\label{params_app1}
\end{table}

\section{Application II: Rubin bright-star masks over the entire survey footprint}
\label{sec:examplerubin}
As a second application of Skykatana, we demonstrate how to construct bright-star exclusion masks within the notebook aspect of the Rubin Science Platform environment by querying and pixelating Gaia stars on-demand. This example (albeit with a much simpler input stage) is covered in more detail in  \href{https://github.com/samotracio/skykatana/blob/main/notebooks/quick_example_rubin.ipynb}{this notebook}. In order to extract the various regions defined in the Rubin baseline footprint, please follow \href{https://github.com/lsst/rubin_sim_notebooks/blob/main/scheduler/3-MDP_surveys.ipynb}{this notebook} and the steps \href{https://skykatana.readthedocs.io/en/latest/hspbasic/}{here} to ingest it as a pipeline stage.

\subsection{Inputs and the RSP platform context}
First, we have to define a search area in which to look for stars. While it can be any stage or boolean map, in Rubin it will most likely be derived from some property map (e.g. coverage in a given band or total exposure time). We choose the nominal Wide-Fast-Deep (WFD) region, as planned with the baseline 10yr observing strategy. The WFD, which comprises more than 17600 deg$^2$, is expected to consume $\sim$80\% of the total survey time. It includes both low-dust extinction areas that are useful for extragalactic science, and higher stellar density areas, adequate for galactic research.

Assuming that \texttt{lowdust} is the WFD map of the r-band, we ingest it as a pipeline stage with 

\begin{lstlisting}[language=Python, caption={Initialization of a pipeline in Skykatana}]
mkp = SkyMaskPipe()
mkp.lowdustmask = lowdust
\end{lstlisting}

\begin{figure}
    \centering
    \includegraphics[width=1\linewidth]{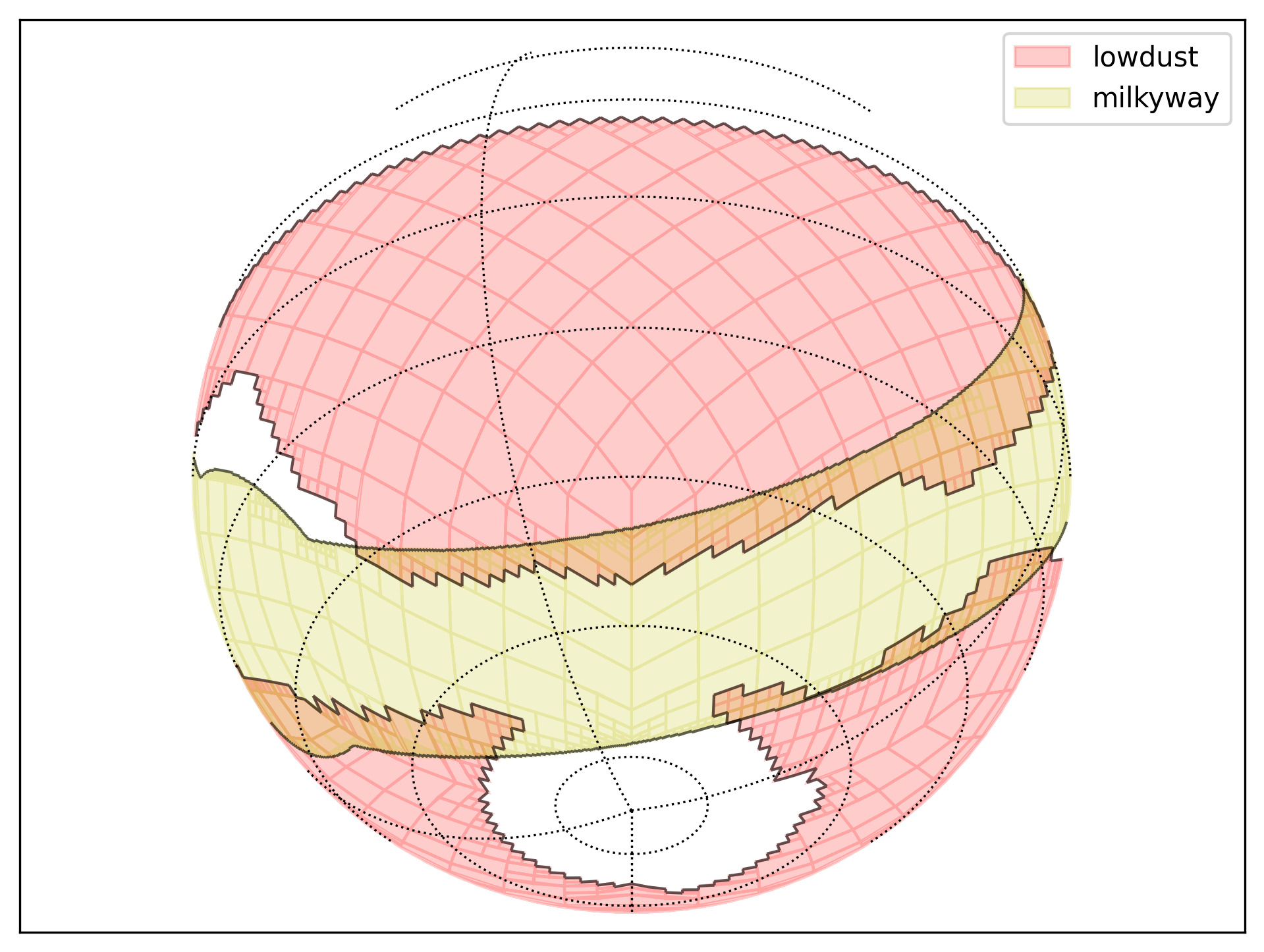}
    \caption{MOC representation of the WFD area selected as a search stage for querying Gaia stars (SIN projection in WCS axes).}
    \label{fig:wfd}
\end{figure}

In Figure \ref{fig:wfd} we plot the WFD MOC using the standard tools available in Skykatana. Second, we need to define the function that assigns radius to circles around Gaia stars, and we choose the relation for the 85\% completeness level in the Rubin r-band explained in Sect. \ref{sec:radius}. In general, this radius function should take an input dataframe containing Gaia stars and add a 'radius' column (in degrees). One possible implementation is listed below.

\begin{lstlisting}[language=Python, caption={Example of the empirical radius–magnitude function from DP1 data.}]
import numpy as np, pandas ad pd

def radfunction(df: pd.DataFrame, **kwargs):
    # Unpack keyword arguments
    mag = 'phot_g_mean_mag'           # Gaia magnitude
    band = kwargs.get('band', 'i')    # Fitted band
    comp = kwargs.get('comp', 0.85)   # Completeness
    mag = 'phot_g_mean_mag'
    if comp==0.85:            
        if band=='i':
            A = 1056.128;   B = -0.245
    # Set radius
    df['radius'] = A*np.exp(B*df[mag])/3600.
\end{lstlisting}

Now we define the star query dictionary. This controls where and which stars will be queried, as well as some performance parameters to keep memory usage under control. This dictionary encapsulates: (1) the search stage defined above, (2) the location of the query catalog, in this case the Gaia endpoint available in RSP, (3) the columns to retrieve, (4) the magnitude limits of the start to be retrieved, (5) the radius function defined above, which must be wrapped in a \texttt{partial} construct, (6) the switch to avoid the Milky Way and various geometrical parameters, and (7) several performance parameters that control the splitting of the search stage MOC.

\begin{lstlisting}[language=Python, caption={Example star query dictrionary.}]
starq = {
    # Stage where we want to retrieve stars
   'search_stage': mkp.lowdustmask, 
   # URL endpoint of HATS catalog, e.g. Gaia
   'cat': "https://data.lsdb.io/hats/gaia_dr3",
   # Columns to retrieve from the catalog
   'columns':['ra','dec','phot_g_mean_mag'],
   # G_band magnitude limits for Gaia stars
   'gaia_gmag_lims':[9,20],
   # Star radius function. Must be wrapped with partial
   'radfunction': partial(radfunction, band='i', comp=0.85),  
   # Avoid Milky Way switch
   'avoid_mw': True,
   # Mikly Way plane zone of exclusion (deg)
   'b0_deg': 15.,
   # Milky Way bulge semimajor axis length (deg)
   'bulge_a_deg': 25.,
   # Milky Way bulge semiminor axis length (deg)
   'bulge_b_deg': 20., 
   # Optional - max area to per single query (deg2)
   'max_area_single': 800.,  
   # Optional - target area of chunks that the search_stage is splitted into
   'target_chunk_area': 800.,
   # Optional - coarse order to perform MOC chunking
   'coarse_order_bfs': 5
}
\end{lstlisting}

\subsection{Building and inspecting the mask}

After the search stage and query dictionary are ready, we should initialize the Dask client. For RSP we choose 3 workers using up to 6 GB each, as the total memory of a regular instance is constrained to 16GB (workers rarely consume their entire memory allocation at the same time).

\begin{lstlisting}[language=Python, caption={Initializing dask cluster in RSP and running the mask building method}]
from dask.distributed import Client
client = Client(n_workers=3, threads_per_worker=1, memory_limit="6GiB")
mkp.build_star_mask_online(starq=starq, order_sparse=15, order_cov=5, n_threads=1, save_stars=True);
\end{lstlisting}

The code proceeds as described in Sect. \ref{sec:starmaskbuilder}: it converts the MOC of the search region into BF chunks, adds a pixel around borders, queries the Gaia HATS endpoint for each chunk, assigns radii using the DP1 calibrated function, and pixelizes each exclusion disc into a bit-packed boolean map. The partial maps are then streamed and merged into a single stage of (customizable) name \texttt{starmask}. The result is a mask consisting of $\sim$2 billion pixels of order 15 (6.4'' pixel size) produced by pixelating about 110 million stars. The mask due to stars has a total area of ~6480 deg$^2$ that with the above configuration is computed in around 68 minutes in the RSP. Key parameters used in this application are listed in Table \ref{params_app2}.

\begin{figure}[htbp]
    \centering
    \includegraphics[width=0.5\textwidth]{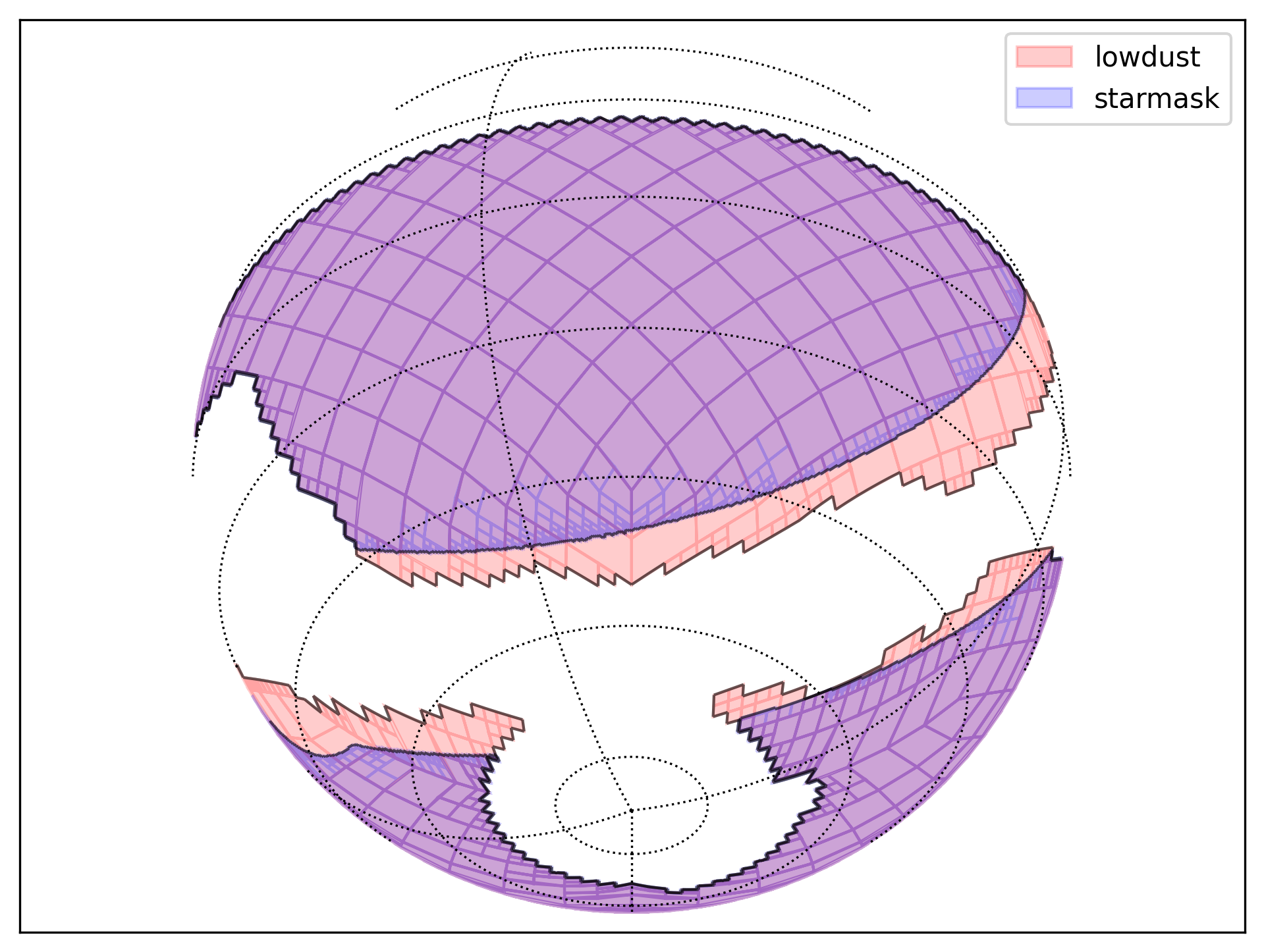}\\[0em] 
    \includegraphics[width=0.5\textwidth]{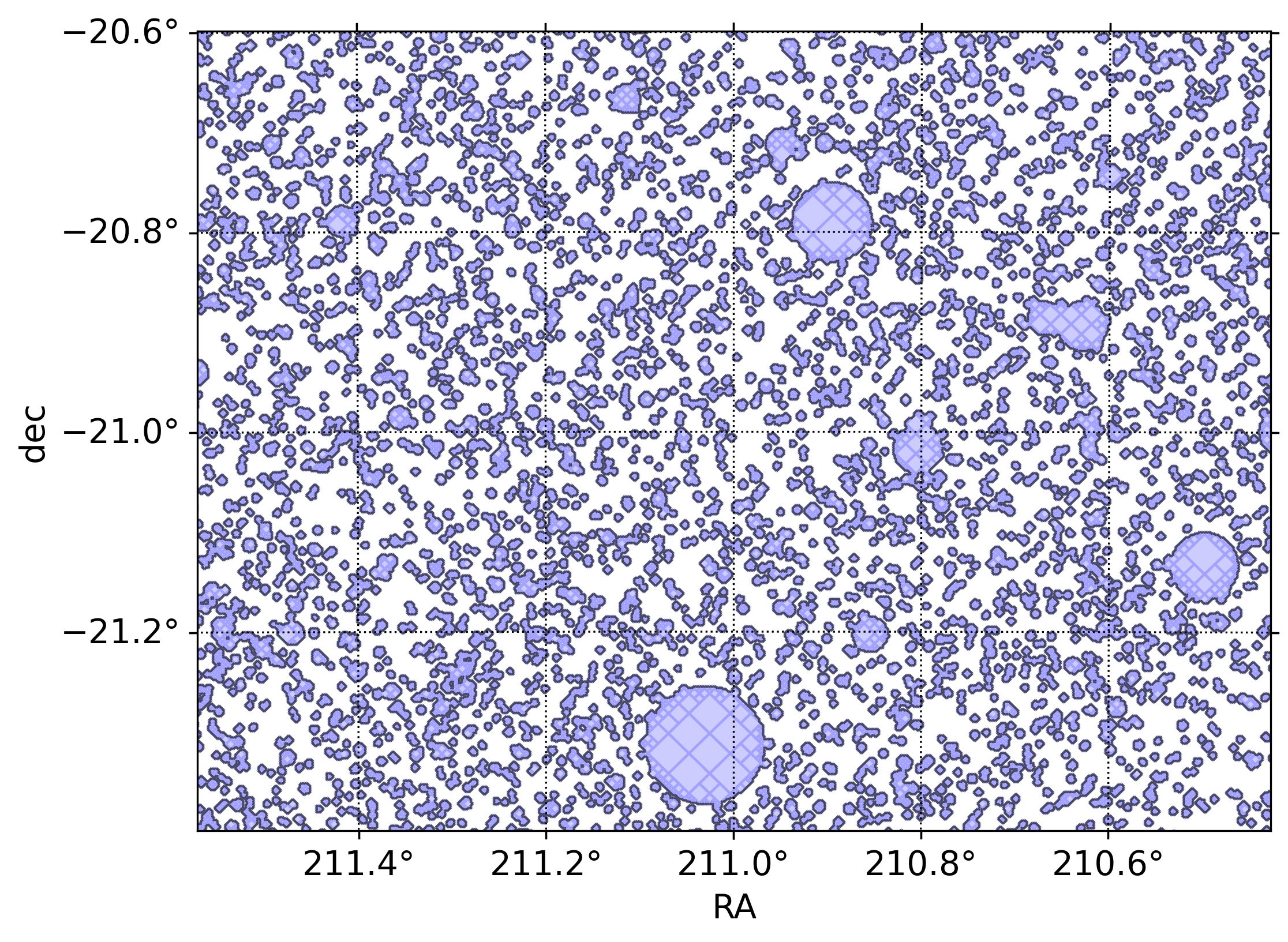}
    \caption{Top: bright-star mask constructed over the WFD region as planned for the 10-yr survey by the Rubin Observatory. Bottom: zoom of the star mask, showing the circles pixelated around Gaia stars of different magnitudes.}
    \label{fig:starmask}
\end{figure}

In Figure \ref{fig:starmask} we show the MOC of the star mask generated from Gaia stars. The effect of the Milky Way avoidance mask is clearly visible, and zooming in on a small region highlights the numerous circles that will perforate the area. Depending on the scientific purposes, the user can customize the mask to include stars in a specific magnitude range or completely change the radius-magnitude relation. 

\begin{figure}[htbp]
    \centering
    \includegraphics[width=0.48\textwidth]{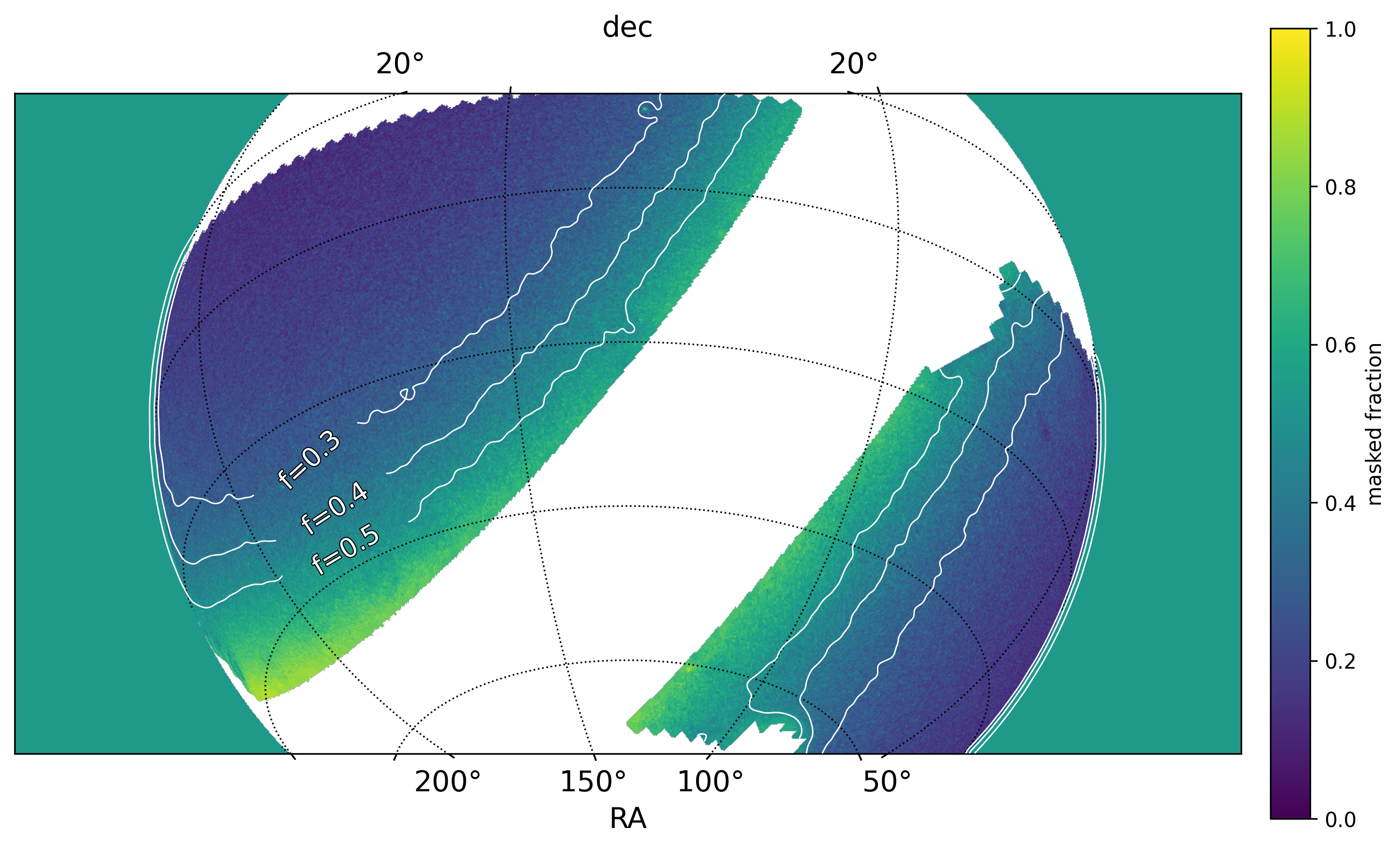}\\[0em] 
    \includegraphics[width=0.48\textwidth]{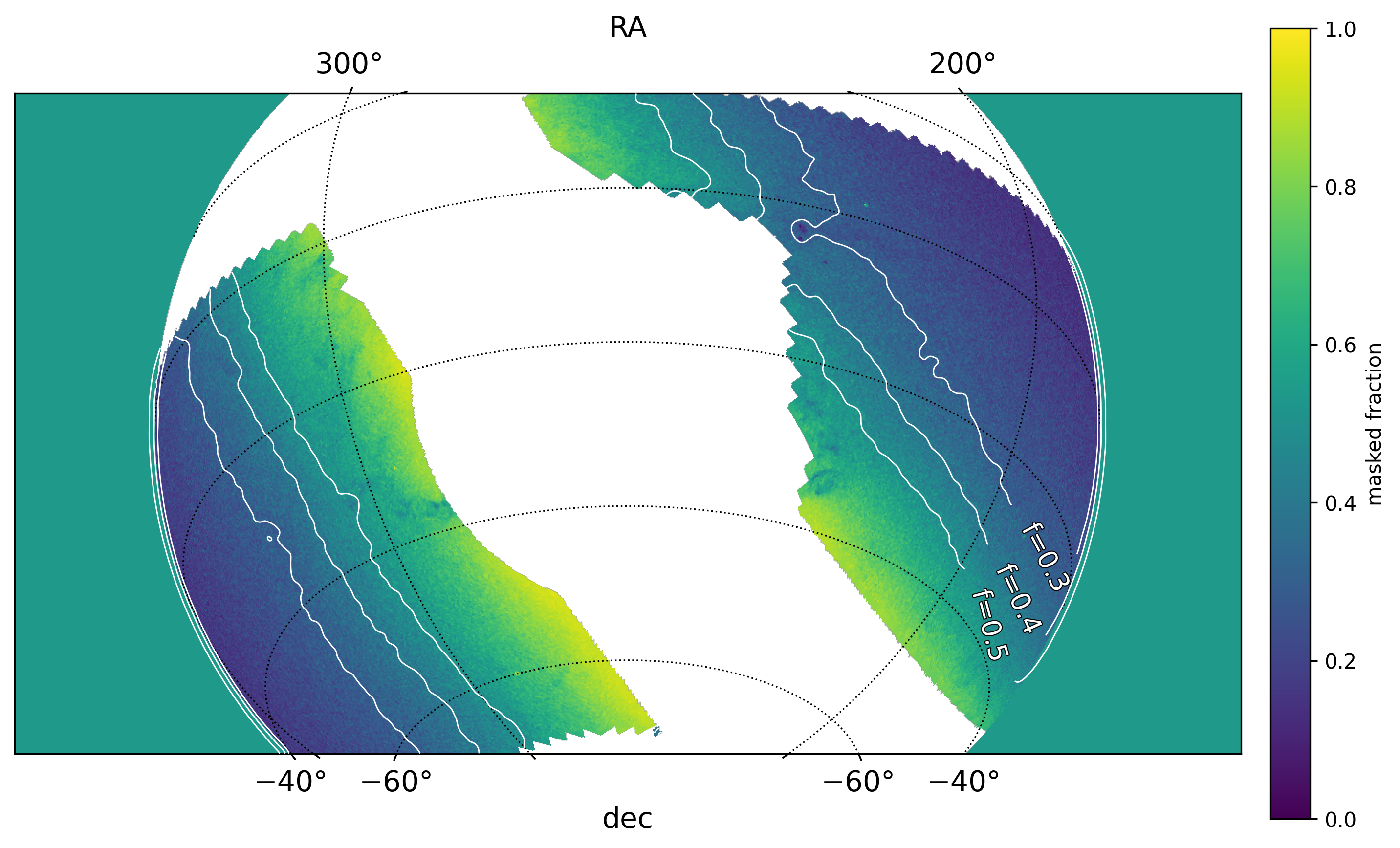}\\[0em]
    \includegraphics[width=0.48\textwidth]{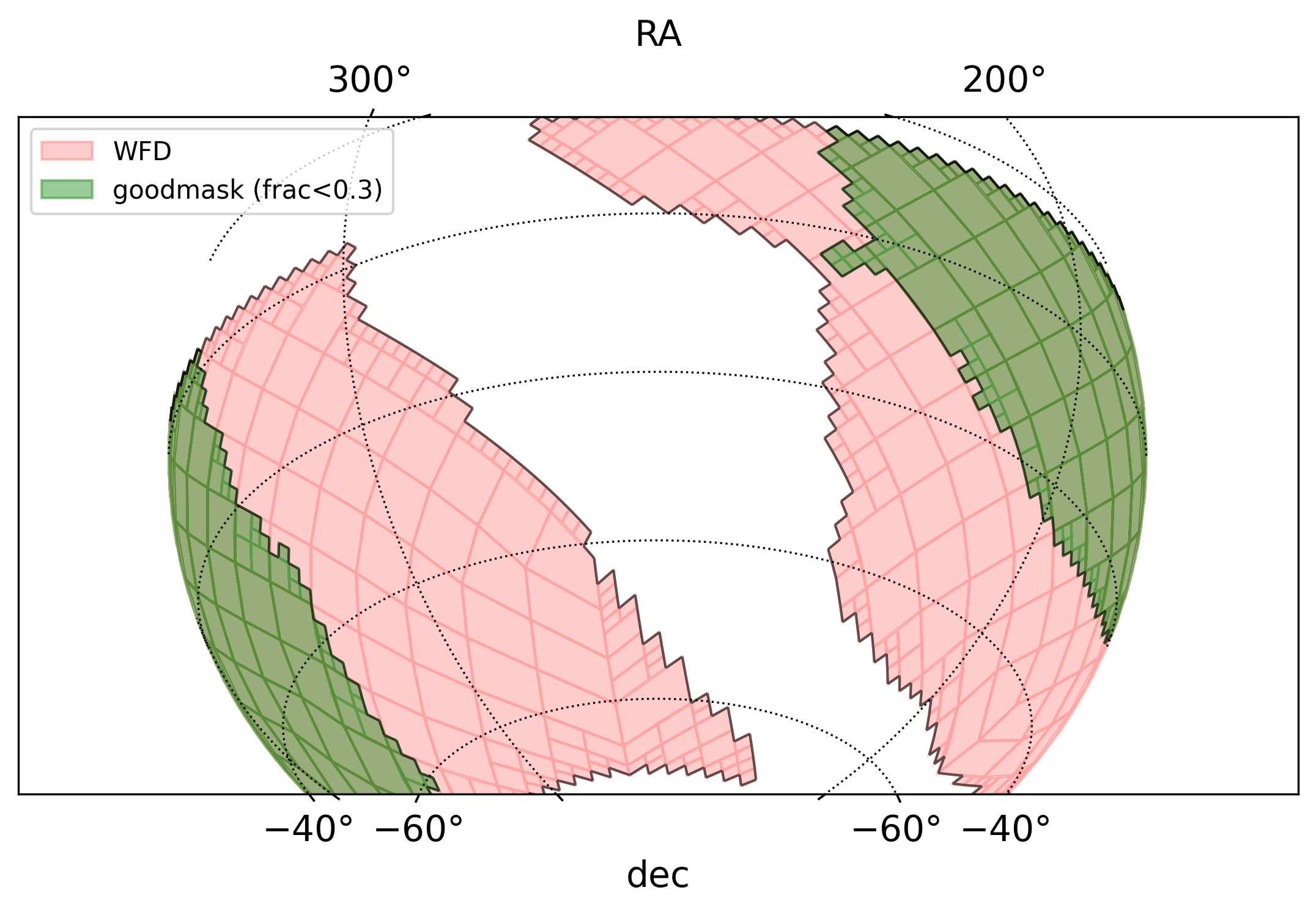}\\[0em]
    \caption{Top and middle: fractional area map masked by stars over the WFD footprint. Yellow regions are heavily affected, while darker regions lose many fewer pixels due to bright stars. Contours enclose zones that would have up to 50\% (yellow), 40\% (green) and 30\% (dark blue) of its area masked by stars. Bottom: fractional area low}
    \label{fig:starfrac}
\end{figure}

When investigating the effect of stars and other sources, a useful statistical measure is the fractional area masked out, i.e. the fraction of sparse pixels masked by those objects at a given order. In dense regions near the galactic plane, the number of pixels affected by stars can be so high that this fraction can approach one. In contrast, zones away from crowded stellar fields typically display fractions of 0.1-0.3 for most choices of the radius-magnitude relationship. By construction, such a fraction will be artificially low for pixels around the edge of a mask. Since this can lead to odd artifacts when performing boolean operations, Skykatana implements an optional switch that replaces the values of the edge pixel by their local harmonic mean. In this way, the code can automatically calculate the fractional map of any stage, generate its 2D image representation, and overlay it over a given WCS axes with optional contour levels. In Figure \ref{fig:starfrac} we show the WFD fractional area map of order 8, that is lost after masking Gaia stars. As expected, regions at low galactic latitudes are heavily affected. The contours enclose zones that would have up to 50\% (yellow), 40\% (green), and 30\% (dark blue) of their area masked by stars. 

With the tools available in Skykatana, it is straightforward to create a mask that tolerates losing, for example, up to 30\% of its area in stars. This \texttt{goodmask} is shown in the bottom panel of Figure \ref{fig:starfrac}, and was created with the following code.

\begin{lstlisting}[language=Python, caption={Thresholding a fractional area map.}]
# Create fractional map and a temporary stage
frac = SkyMaskPipe.frac_area_map(mkp.starmask, order_frac=5, avg_edges=True)
mkp.build_prop_mask(prop_maps=[frac],thresholds=[0.3],comparisons=['lt'],output_stage='tmpmask', order_sparse=5, order_cov=3)
# Refine temporary stage to follow more closely the input WFD area, subtracting the MW vetto mask
mkp.combine(positive=[('lowdustmask','tmpmask')], negative=['mwmask'], order_out=5, order_cov=3, output_stage='goodmask')
\end{lstlisting}

\begin{table}[hb]
\begin{adjustbox}{max width=\linewidth}
\begin{tabular}{lc}
\hline
Parameter & Value \\
\hline
Search stage & WFD low-dust footprint \\
HEALPix sparse order & 15 \\
HEALPix coverage order & 5 \\
Gaia magnitude range & $9 < G < 20$ \\
Chunk target area & 800 deg$^2$ \\
Coarse BFS order & 5 \\
Milky Way exclusion & $|b| < 15^\circ$ + elliptical bulge mask \\
Radius function & Empirical DP1 $R(G)$ relation (85\% completeness) \\
Bit-packed storage & Enabled \\
Dask workers / memory & 3 workers, 6 GB each (max) \\
\hline
\end{tabular}
\end{adjustbox}
\caption{Key configuration parameters for the Rubin bright-star mask demonstration (Application II).}
\label{params_app2}
\end{table}

The on-demand star-mask construction described here leverages LSDB/HATS catalogs hosted within the Rubin Science Platform. However, the workflow is not RSP-specific. Users can reproduce the analysis outside of RSP by providing any local or remote catalog accessible via Dask, and just pointing the query interface to the HATS catalog folder. For environments without LSDB, Skykatana also supports building star masks from user-supplied catalogs in a variety of formats via the \texttt{build\_circ\_mask()}  method, enabling full reproducibility on local computing systems. 

In Table~\ref{tab:performance} we summarize the computational requirements for two representative Skykatana applications. The reported figures correspond to end-to-end wall-clock execution times, and peak memory usage. We note that the catalog-query component of the pipeline is executed on the Rubin Science Platform, which is deployed on a shared cloud infrastructure. As a result, absolute query times can vary between runs depending on system load, caching effects, and HATS endpoint location.

\begin{table}[ht]
\begingroup
\begin{adjustbox}{max width=\linewidth}
\begin{tabular}{lccccc}
\hline
Application & Area (deg$^2$) &  Order & $t_{wall}$ & $t_{pix}$ & $m_{peak}$ \\
\hline
HSC--WISE (local) & 69 & 13--15 & 6 s & --- & --- \\
Rubin WFD (RSP)& 17645 & 15 & 65 m & 53 m & 2.3 GB \\
\hline
\end{tabular}
\end{adjustbox}
\endgroup
\caption{Summary of computational requirements for representative Skykatana applications. The columns correspond to the search stage area, wall clock time, pixelization time (single thread), and peak memory usage. The RSP example was run with 3 Dask workers and 1 thread per worker. For the HSC--WISE application we quote only the end-to-end runtime (cell 3 of the notebook), as the total execution time is sufficiently short that a robust separation between I/O and pixelization costs is not meaningful.}
\label{tab:performance}
\end{table}


\section{Discussion}
Skykatana demonstrates that large-scale angular masking can be achieved efficiently through sparse pixelization, streaming I/O, and modular composition. Its design deliberately prioritizes memory efficiency and scalability, enabling users to build multi-billion-pixel masks within the constrained resources typical of shared-notebook and cloud-based environments. The framework bridges the gap between polygon-based approaches such as \texttt{mangle} and modern HEALPix-based pipelines by providing a high-level abstraction that is both intuitive and computationally efficient. By leveraging the dual “coverage/sparse” structure of HealSparse and combining it with MOC-based visualization, Skykatana achieves a balance between geometric precision, scalability, and usability for survey-scale data analysis.

Modern wide-area surveys such as Rubin, Euclid, and Roman increasingly require angular masks at resolutions corresponding to billions of HEALPix pixels, particularly to mitigate stellar contamination, artifacts, and spatially varying completeness. In practice, these masks must be constructed, inspected, and iterated upon within shared, notebook-based, or cloud environments where memory and I/O resources are constrained. Skykatana is explicitly designed for this regime, enabling the construction, manipulation, and interactive visualization of such masks through sparse, bit-packed representations and streaming I/O. A central contribution of the framework is its ability to construct masks directly from distributed catalog queries rather than from pre-materialized full-sky products, by streaming only the relevant sky regions and magnitude ranges through a chunked workflow. Although demonstrated here using the Rubin Science Platform infrastructure, the same design naturally applies to other large-survey analysis ecosystems based on distributed data access.

To represent survey masks on the sphere, we adopt the HEALPix tessellation. Although HEALPix pixels are not identical in shape, all pixels at a given resolution have exactly the same area and form a hierarchical and iso-latitude grid (\citealt{gorski2005}). This equal-area property is particularly relevant for angular masks and clustering analyses, as it avoids artificial density fluctuations introduced by variable pixel areas. The hierarchical structure further enables efficient multi-resolution representations, chunked I/O, and boolean mask algebra, while the iso-latitude sampling facilitates fast indexing and spherical operations. Alternative tessellations such as the Hierarchical Triangular Mesh (HTM; \citealt{kunszt2001}) provide more regular geometric elements and fast indexing, but do not guarantee equal-area pixels and are less commonly adopted in modern survey masking frameworks. For these reasons, HEALPix has become the de facto standard in cosmological surveys (e.g., Planck, Gaia, Rubin, Euclid), and is adopted here for statistical robustness and interoperability.

Compared to the HSC-SSP bright-star mask approach, Skykatana generalizes the concept from static survey products to a reproducible procedural pipeline. The resulting maps are automatically extensible to arbitrary sky regions, filters, completeness thresholds, and user-defined magnitude--radius relations. Relative to \texttt{mocpy}, which focuses on set operations over multi-order coverage maps, Skykatana operates at substantially higher pixel orders and retains explicit boolean semantics at the sparse-pixel level, enabling fine-grained control over mask geometry.

Several simplifying assumptions remain to be addressed. Circular exclusion zones provide a practical approximation to stellar halos but cannot fully capture the complex anisotropy of diffraction spikes, ghosts, or diffuse scattered light. In addition, boolean masking inherently treats all affected pixels equally, whereas partial weighting schemes could preserve useful information in regions of low contamination. Extending Skykatana to support non-boolean (integer or floating-point) maps, such as weight, depth, or completeness fields, would broaden its scope to probabilistic selection functions and variable-coverage analyzes.

Finally, the ECDFS field observed by Rubin is, by construction, a region largely devoid of very bright stars, whereas other fields included in DP1 have not yet been observed to full survey depth. As a consequence, the radius--magnitude relation derived in Section~\ref{sec:radius} is not well constrained for stars brighter than approximately 8--8.5 mag. As the Rubin survey progresses and deeper data become available over a wider range of environments, this relation will be recalibrated to better characterize the impact of bright stars on survey completeness.

\section{Summary and conclusions}
We have presented an open-source, scalable framework for constructing, combining, and visualizing survey-scale sky masks. Built on the HEALPix and HealSparse infrastructures, Skykatana provides a unified pipeline architecture that enables the creation of complex boolean maps, footprints, bright-star masks, and quality threshold maps using minimal memory resources. Two representative applications were demonstrated:

\begin{itemize}
    \item An HSC–WISE composite mask, combining footprint, patch quality, and artifact stages.

    \item Rubin bright-star masks, generated on demand in the Rubin Science Platform by querying Gaia DR3 stars through LSDB/HATS and applying empirically calibrated radius–magnitude relations from Rubin DP1 data in the ECDFS field.
\end{itemize}

These examples show that arcsecond resolution masks of up to billions of pixels can be generated in a few dozen minutes with modest computing resources while maintaining full reproducibility and traceability of parameters. Skykatana visualization methods provide immediate diagnostic power for assessing the spatial impact of masking decisions. Looking ahead, Skykatana can serve as a foundational layer for next-generation survey pipelines that demand spatial reproducibility at scale. By extending the current boolean representation to include weighted and continuous fields, the framework could integrate seamlessly with probabilistic catalog analyzes, joint likelihood mapping, and end-to-end selection function modeling for Rubin and beyond.

In this work, the empirical demonstrations are intentionally focused on validating the performance, scalability, and reproducibility of the Skykatana framework using real survey data. The bright-star masking examples are presented as concrete use cases illustrating how catalog-driven masks can be constructed and applied at scale. Broader calibration validity, systematic field-to-field comparisons, and more complex masking schemes, such as anisotropic or environment-dependent masks, are not claimed here as validated results; rather, they are natural extensions enabled by the framework as larger and more diverse data sets become available. We are currently extending Skykatana to support floating-point (weighted) stellar masks that encode a probabilistic completeness correction rather than strict binary exclusion. Such mechanism can potentially retain a large fraction of area/sources that would otherwise get excluded, and even improve noise properties. A dedicated follow-up paper will present the method and quantify its impact on the large-scale structure (e.g. clustering measurements).

\section*{Acknowledgements}
The authors thank the referees for useful and constructive comments that improved this manuscript.
This project was made possible through a collaboration with the LINCC Frameworks software engineering team as part of the Frameworks Incubator Program, which is supported by the LSST Discovery Alliance (LSST-DA). 

M.D. work was supported by the Preparing for Astrophysics with LSST Program, funded by the Heising Simons Foundation through grant 2021-2975, and administered by Las Cumbres Observatory.

This research has also made extensive use of the following software: Astropy \citep{astropy}, Matplotlib \citep{matplotlib}, Pandas \citep{pandas}, Seaborn \citep{seaborn}, Healpy \citep{healpy}, SciPy \citep{scipy}, NumPy \citep{numpy}, HEALSparse \citep{hsppypi},  MOCpy \citep{boch2019},   HATS/LSDB schemas \citep{caplar2025}, The Jupyter Notebook \citep{jupyter-notebook}, and TOPCAT \citep{topcat}.

This work has made use of data from the European Space Agency (ESA) mission
{\it Gaia} (\url{https://www.cosmos.esa.int/gaia}), processed by the {\it Gaia}
Data Processing and Analysis Consortium (DPAC, \url{https://www.cosmos.esa.int/web/gaia/dpac/consortium}). Funding for the DPAC has been provided by national institutions, in particular the institutions participating in the {\it Gaia} Multilateral Agreement. We use the HATS/LSDB Gaia-DR3 Catalogs hosted at \url{https://data.lsdb.io/}.

The Hyper Suprime-Cam (HSC) collaboration includes the astronomical communities of Japan and Taiwan, and Princeton University. The HSC instrumentation and software were developed by the National Astronomical Observatory of Japan (NAOJ), the Kavli Institute for the Physics and Mathematics of the Universe (Kavli IPMU), the University of Tokyo, the High Energy Accelerator Research Organization (KEK), the Academia Sinica Institute for Astronomy and Astrophysics in Taiwan (ASIAA), and Princeton University. Funding was contributed by the FIRST program from the Japanese Cabinet Office, the Ministry of Education, Culture, Sports, Science and Technology (MEXT), the Japan Society for the Promotion of Science (JSPS), Japan Science and Technology Agency (JST), the Toray Science Foundation, NAOJ, Kavli IPMU, KEK, ASIAA, and Princeton University. 

This paper makes use of software developed for Vera C. Rubin Observatory. We thank the Rubin Observatory for making their code available as free software at http://pipelines.lsst.io/.

This paper is based on data collected at the Subaru Telescope and retrieved from the HSC data archive system, which is operated by the Subaru Telescope and Astronomy Data Center (ADC) at NAOJ. Data analysis was in part carried out with the cooperation of Center for Computational Astrophysics (CfCA), NAOJ. We are honored and grateful for the opportunity of observing the Universe from Maunakea, which has the cultural, historical and natural significance in Hawaii. 

This publication makes use of data products from the Wide-field Infrared Survey Explorer, which is a joint project of the University of California, Los Angeles, and the Jet Propulsion Laboratory/California Institute of Technology, funded by the National Aeronautics and Space Administration.

This research use data from the NSF-DOE Vera C. Rubin Observatory (2025); Legacy Survey of Space and Time Data Preview 1 \citep{10.71929/rubin/2570308} and materials from NSF-DOE Vera C. Rubin Observatory (2025); Rubin Observatory LSST Tutorials
\citep{10.11578/rubin/dc.20250909.20}.

This material is based on work supported in part by the National Science Foundation through Cooperative Agreements AST-1258333 and AST-2241526 and Cooperative Support Agreements AST-1202910 and 2211468 managed by the Association of Universities for Research in Astronomy (AURA), and the Department of Energy under Contract No. DE-AC02-76SF00515 with the SLAC National Accelerator Laboratory managed by Stanford University. Additional Rubin Observatory funding comes from private donations, grants to universities, and in-kind support from LSST-DA Institutional Members.

\appendix

\section{Mask Products Release}
Using the magnitude--radius relations calibrated as described in Section \ref{sec:radius} for each Rubin band, we run Skykatana algorithms in RSP to build masks at order 15 (pixel size $\sim~6.4''$) for the WFD, NES, SCP and LMC-SMC areas (see \href{https://github.com/lsst/rubin_sim_notebooks/blob/main/scheduler/3-MDP_surveys.ipynb}{here} for their definition). We also include the LSST Survey Validation area. To facilitate using the masks for different purposes, we also split into two star magnitude ranges, $9<G_{\mathrm{Gaia}}\leq18$ and $18<G_{\mathrm{Gaia}}\leq20$.  Together, they comprise about 22 billion pixels covering a large fraction of the southern sky (note that some of these pixels overlap per band, between non-halo/halo versions, and between magnitude cuts). Table \ref{maskrelease} shows their main statistical properties, and mask files are publicly available \href{https://osf.io/r5vw6}{here}. We plan to offer a user-configurable angular mask service through the Argentine Rubin Independent Data Access Center in the near future. 


\begin{table}[htbp]
\centering
\begin{adjustbox}{max width=\linewidth}
\begin{tabular}{l r r r} 
\hline
Band [mag. range($G_{\mathrm{Gaia}}$)] & $n_\mathrm{pix}$ & area~$[\mathrm{deg^2}]$ & $\langle f_{85} \rangle$ \\ 
\hline
Wide-Fast-Deep (WFD) \\
g [9,18] & 1920962337 & 6150.2 & 0.38  \\ 
g [9,18] halo & 2993748958 & 9564.9 & 0.60  \\ 
g [18,20]& 939286935 & 3007.3 & 0.19  \\ 
r [9,18] & 1807071944 & 5785.6 & 0.36  \\ 
r [9,18] halo & 2789266836 & 8930.3 & 0.56  \\
r [18,20]& 891382011 & 2853.9 & 0.18  \\ 
i [9,18] & 1622223085 & 5193.8 & 0.32  \\ 
i [9,18] halo & 2070959961 & 6630.5 & 0.41  \\ 
i [18,20]& 784875062 & 2512.9 & 0.16  \\ 
North-Ecliptic-Survey (NES) \\ 
g [9,18] & 219412803 & 702.5 & 0.32  \\ 
g [9,18] halo & 374344055 & 1198.5 & 0.55  \\ 
g [18,20]& 81253964 & 260.1 & 0.12  \\ 
r [9,18] & 204556039 & 654.9 & 0.30  \\ 
r [9,18] halo & 345013604 & 1104.6 & 0.51  \\ 
r [18,20]& 76453899 & 244.8 & 0.11  \\ 
i [9,18] & 181223224 & 580.2 & 0.27  \\ 
i [9,18] halo & 243075730 & 778.2 & 0.36  \\
i [18,20]& 66030155 & 211.4 & 0.10  \\ 
South-Celestial-Pole (SCP) \\
g [9,18] & 50719032 & 482.5 & 0.61  \\ 
g [9,18] halo & 199321725 & 638.2 & 0.81  \\
g [18,20]& 94540224 & 302.7 & 0.38  \\ 
r [9,18] & 143620898 & 459.8 & 0.58  \\ 
r [9,18] halo & 190013019 & 608.4 & 0.77  \\ 
r [18,20]& 90213626 & 288.8 & 0.37  \\ 
i [9,18] & 131299570 & 420.4 & 0.53  \\ 
i [9,18] halo & 155560680 & 498.0 & 0.63  \\ 
i [18,20]& 80394581 & 257.4 & 0.33  \\ 
Large/small Magellanic Clouds (LMC-SMC) \\
g [9,18] & 52478738 & 168.0 & 0.51  \\ 
g [9,18] halo & 71023695 & 227.4 & 0.70  \\ 
g [18,20]& 47417536 & 151.8 & 0.46  \\ 
r [9,18] & 50091914 & 160.4 & 0.49 \\ 
r [9,18] halo & 67144235 & 215.0 & 0.66  \\ 
r [18,20]& 46311751 & 148.3 & 0.45  \\ 
i [9,18] & 45983457 & 147.2 & 0.45  \\ 
i [9,18] halo & 53845209 & 172.4 & 0.53  \\ 
i [18,20]& 43701154 & 139.9 & 0.43  \\ 
Survey Validation (SV) \\
g [9,18] & 275883539 & 883.3 & 0.43  \\ 
g [9,18] halo & 407561897 & 1304.9 & 0.63  \\ 
g [18,20]& 158862842 & 508.6 & 0.25  \\ 
r [9,18] & 311600281 & 997.6 & 0.37 \\ 
r [9,18] halo & 468446865 & 1499.8 & 0.56  \\ 
r [18,20]& 175214019 & 561.0 & 0.21  \\ 
i [9,18] & 348563503 & 1116.0 & 0.32  \\ 
i [9,18] halo & 438402299 & 1403.6 & 0.31  \\ 
i [18,20]& 191628364 & 613.5 & 0.18  \\ 
Extended Chandra Deep Field South (ECDFS) \\
g [9,18] & 230598 & 0.74 & 0.21  \\ 
g [9,18] halo & 479713 & 1.54 & 0.44  \\ 
g [18,20]& 65558 & 0.21 & 0.06  \\ 
r [9,18] & 212643 & 0.68 & 0.20 \\ 
r [9,18] halo & 436120 & 1.40 & 0.40  \\ 
r [18,20]& 61476 & 0.2 & 0.06  \\ 
i [9,18] & 184421 & 0.59 & 0.17  \\ 
i [9,18] halo & 272226 & 0.87 & 0.25  \\ 
i [18,20]& 51984 & 0.17 & 0.05  \\ 
\hline
\end{tabular}
\end{adjustbox}
\caption{Masks for the baseline 10-yr Rubin footprint for each band, and split into two star magnitude ranges, $9<G_{\mathrm{Gaia}}\leq18$ and $18<G_{\mathrm{Gaia}}\leq20$. The $area$ column refers to the area of pixels masked by stars. $\langle f_{85} \rangle$ is the mean fraction of the corresponding targeted LSST area, that is masked by stars at the 85\% level.}
\label{maskrelease}
\end{table}

\begin{figure}[htbp]
    \centering
    \includegraphics[width=1\linewidth]{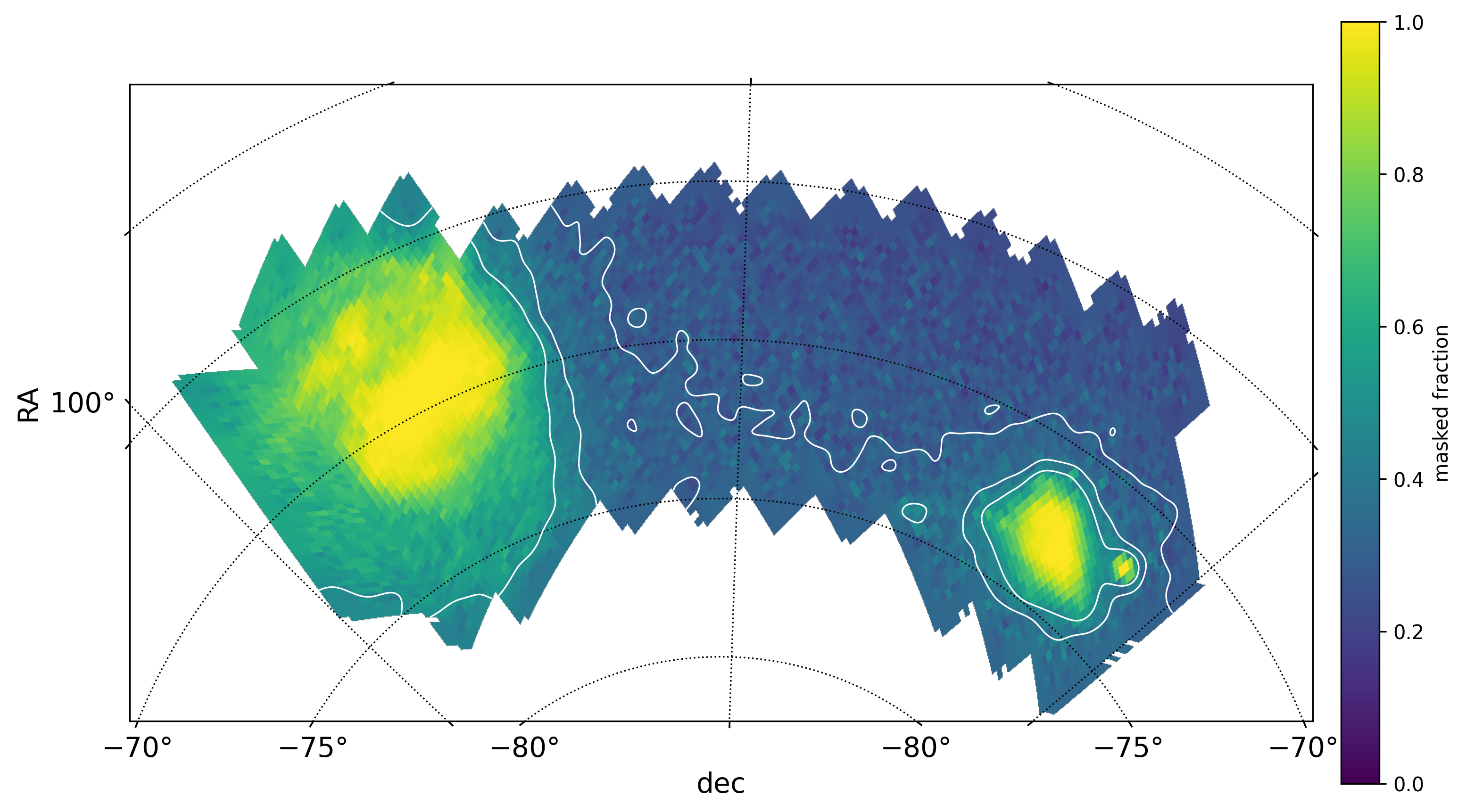}
    \caption{Fractional star map in the i-band for the LMC-SMC region defined by LSST. Both Magellanic clouds are clearly visible. Contours are plotted at 0.3, 0,4 and 0.5 levels.}
    \label{fig:lmcmask}
\end{figure}

\begin{figure}[htbp]
    \centering
    \includegraphics[width=1\linewidth]{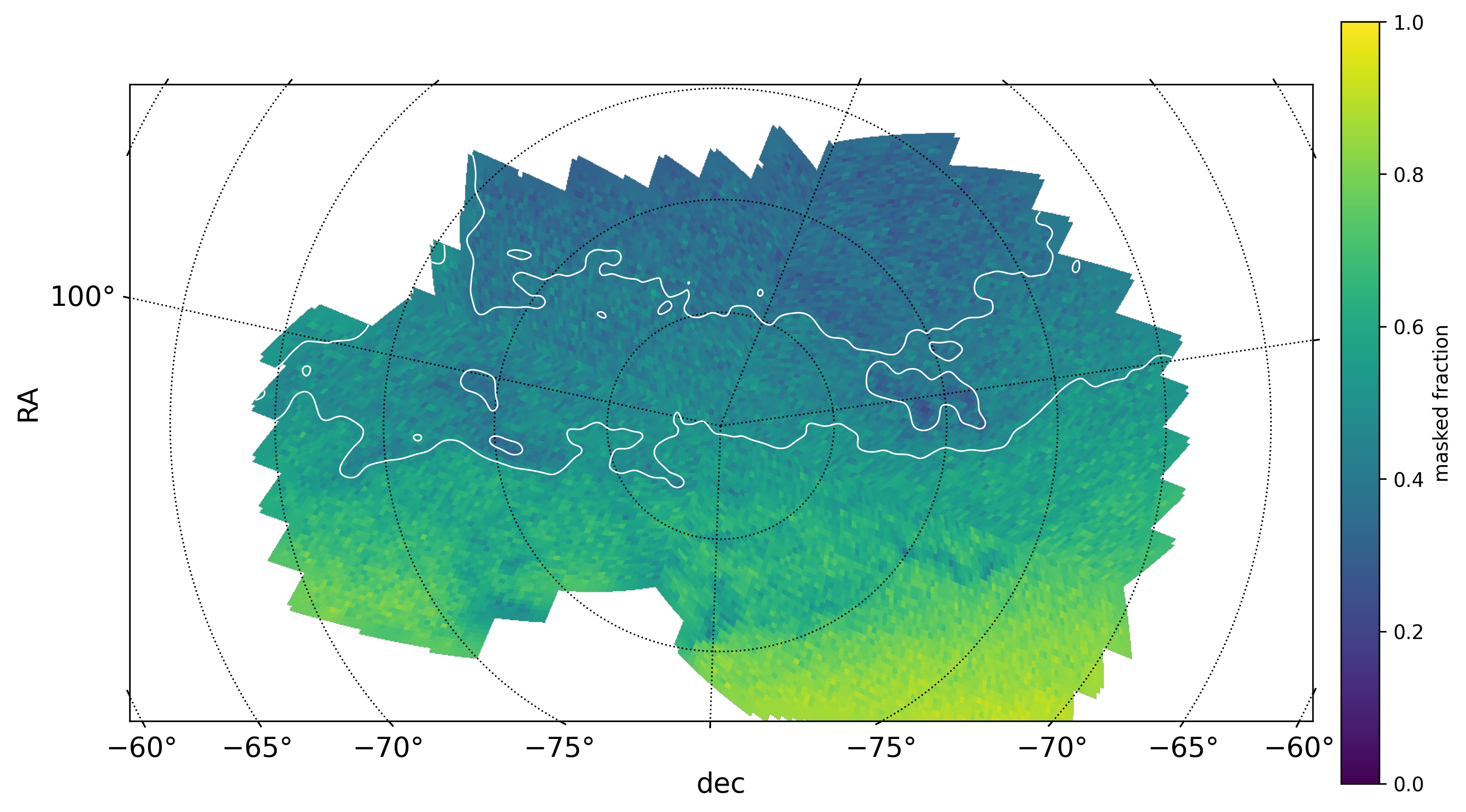}
    \caption{Fractional star map in the i-band for the South Celestial Pole region defined by LSST. Contours are plotted at 0.4 and 0.5 levels.}
    \label{fig:scpmask}
\end{figure}

\begin{figure}[htbp]
    \centering
    \includegraphics[width=1\linewidth]{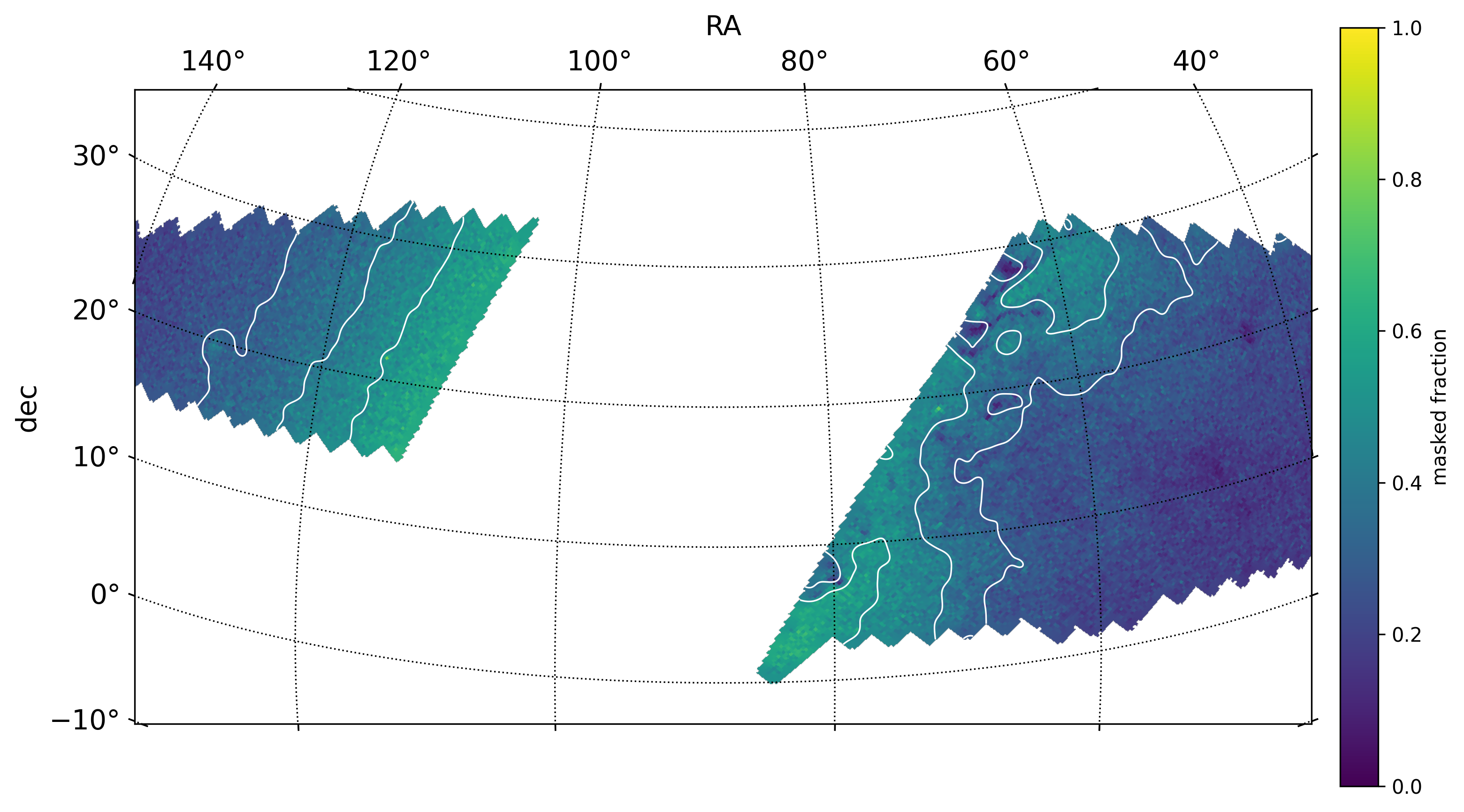}
    \caption{Fractional star map in the i-band for the North Ecliptic Survey region defined by LSST. Contours are plotted at 0.3, 0.4 and 0.5 levels.}
    \label{fig:nesmask}
\end{figure}

\begin{figure}[htbp]
    \centering
    \includegraphics[width=1\linewidth]{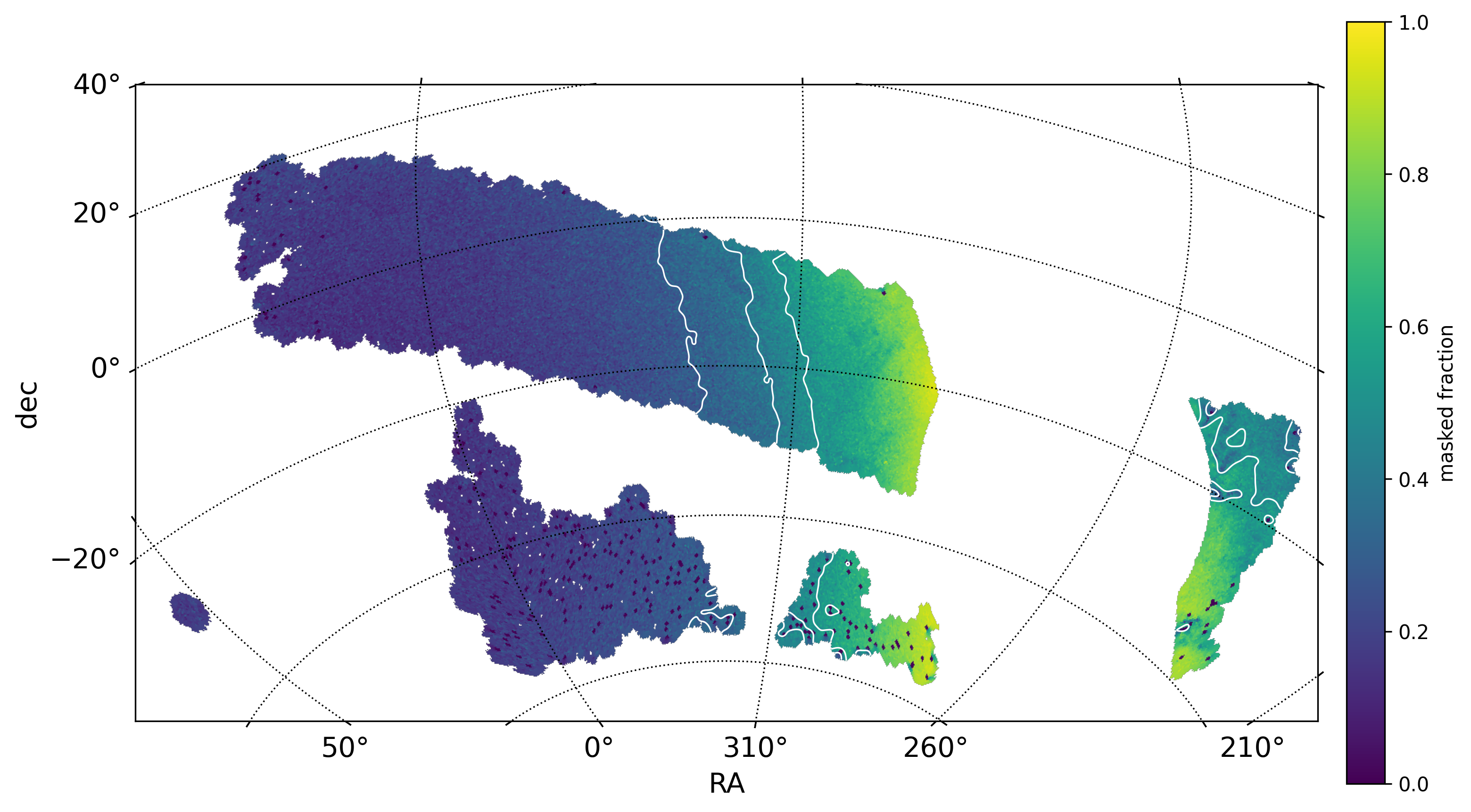}
    \caption{Fractional star map in the i-band for the Survey Validation region defined by LSST (Aitoff projection). Contours are plotted at 0.3, 0.4 and 0.5 levels.}
    \label{fig:svmask}
\end{figure}

\begin{figure}[htbp]
    \centering
    \includegraphics[width=1\linewidth]{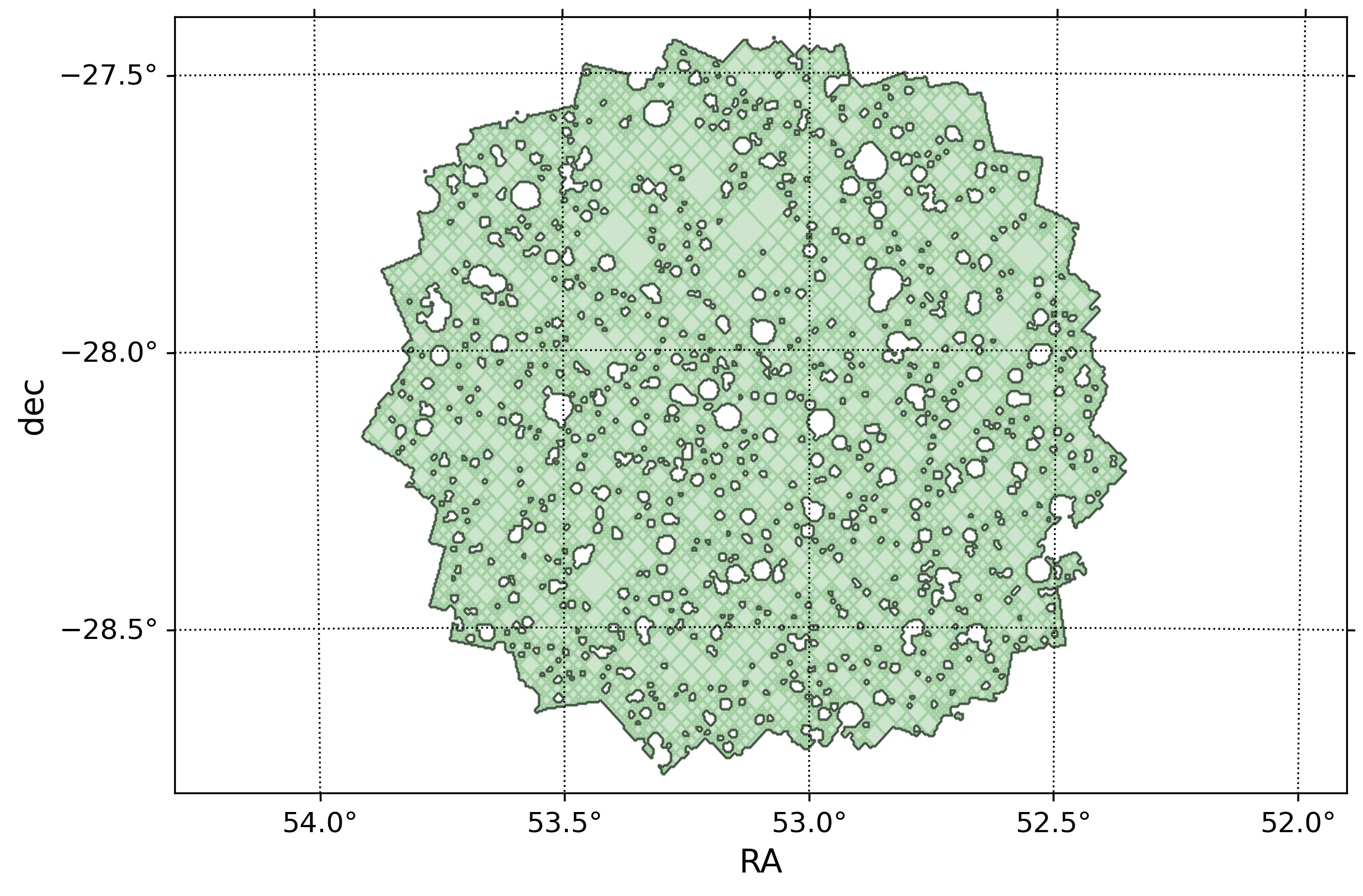}
    \caption{Mask at order 15 (i-band, 85\% level) for the ECDFS region defined by LSST, after subtracting bright stars with magnitudes $9<G_{\mathrm{Gaia}}\leq18$. The bright star halo is not considered.}
    \label{fig:ecdfs}
\end{figure}


\bibliographystyle{elsarticle-harv} 
\bibliography{example}






\end{document}